\documentclass[pra,twocolumn,superscriptaddress,
preprintnumbers,amsmath,amssymb,floatfix]{revtex4-2}
\usepackage{enumerate}
\usepackage{kotex}
\usepackage{graphicx}
\usepackage{float}
\usepackage{amsthm}
\usepackage{tensor}
\usepackage{color}
\usepackage[all]{xy}
\usepackage{tikz}
\usepackage{dsfont}
\usepackage{times,txfonts}
\usetikzlibrary{positioning}
\usepackage{braket}
\usepackage{overpic}
\usepackage{multirow}

\theoremstyle{definition}

\begin{document}




\title{Entanglement recovery by reversing the effect of noise in quantum repeater}




\author{Sewon Jeong}
{\affiliation{Center for Quantum Technology, Korea Institute of Science and Technology (KIST), Seoul, 02792, Korea}
\affiliation{Department of Physics, Ewha Womans University, Seoul, 03760, Korea}

\author{Shrobona Bagchi}
\affiliation{Center for Quantum Technology, Korea Institute of Science and Technology (KIST), Seoul, 02792, Korea}
\affiliation{Quantum Universe Center, Korea Institute for Advanced Study, Seoul, 02455, Korea}

\author{Jaehak Lee}
\affiliation{Center for Quantum Technology, Korea Institute of Science and Technology (KIST), Seoul, 02792, Korea}
\affiliation{Division of Quantum Information, KIST School, Korea University of Science and Technology, Seoul, 02792 Korea}

\author{Hyang-Tag Lim}
\affiliation{Center for Quantum Technology, Korea Institute of Science and Technology (KIST), Seoul, 02792, Korea}
\affiliation{Division of Quantum Information, KIST School, Korea University of Science and Technology, Seoul, 02792 Korea}

\author{Yong-Su Kim}
\affiliation{Center for Quantum Technology, Korea Institute of Science and Technology (KIST), Seoul, 02792, Korea}
\affiliation{Division of Quantum Information, KIST School, Korea University of Science and Technology, Seoul, 02792 Korea}

\author{Taeyoung Choi}
\affiliation{Department of Physics, Ewha Womans University, Seoul, 03760, Korea}

\author{Seung-Woo Lee}
\email{swleego@gmail.com}
\affiliation{Department of Physics, Pohang University of Science and Technology (POSTECH), Pohang 37673, Korea}

\date{\today\\}

\begin{abstract}
We propose a method to directly recover the degree of entanglement distributed by entanglement swapping in the presence of noise. 
Our approach introduces a reversing operation that probabilistically undoes the effect of amplitude damping or photon loss on a single entangled pair, enabling heralded recovery of entanglement. 
We demonstrate that entanglement can be substantially recovered even under strong noise, including parameter regimes where the distributed entanglement would otherwise vanish due to entanglement sudden death. 
We analyze the effectiveness of the protocol in two representative repeater models, i.e.,~two-way and one-way architectures and identify the optimal reversing strategy. 
Due to its heralded and single-copy nature, our protocol is readily compatible with other entanglement recovery techniques such as entanglement purification and distillation. Our work provides a practical and experimentally feasible way toward robust entanglement distribution in current and near-term quantum repeater architectures.
\end{abstract}
\maketitle

\section{Introduction}

Quantum communication offers promising applications in information technologies such as networked quantum computing, quantum sensing, and fundamentally secure communications \cite{Ekert1991, Teleportation,Nielsenbook,Gisin2002,Kimble2008,Lo2014, Duan2001, Gottesman1999, Knill2001, Pirandola2015, DHKim24}. A central requirement for realizing scalable quantum networks is the reliable entanglement distribution over long distances, which typically necessitates quantum repeaters located at intermediate nodes~\cite{Briegel1998,Sangouard2011,Meter2013,Azuma2023,Wehner2018}. A quantum repeater can extend the communication range through entanglement swapping~\cite{Zukowski1993, Pan1998}, where a Bell-state measurement is performed on two qubits from different neighboring nodes to establish entanglement over a longer distance. In principle, by repeating this process across multiple nodes, entanglement distribution over arbitrarily long distances can be achieved.

In realistic scenarios, however, distributed entangled qubits are highly vulnerable to decoherence~\cite{Zurek1991,Breuer2002,Heinz2007,Yu2009}, as well as to inefficiencies in operations and measurements~\cite{Weinfurter94,Calsa2001}.
In particular, noise and photon loss arising in both stationary systems and during transmission can significantly degrade the performance of quantum communication protocols. 
These detrimental effects become more severe as the communication distance increases~\cite{Pirandola2017,QiChao2016,Valivarthi2016,Wzo24,Wzo25}.
Several approaches have been developed to mitigate these problems, including entanglement purification and distillation~\cite{Bennett1996, Deutsch1996, Dur1999,Pan2001,Cheong2007}, advanced Bell-measurement schemes~\cite{Grice2011,SLee13,Ewert2014,Lee2015,Lee2015a} and quantum repeater with error correction encoding~\cite{Jiang2009,Munro10,Munro12,Muralidharan14,Muralidharan16,Azuma15,Ewert16,Lee2019,SML2020}. These methods systematically enhance the quality of entanglement from degraded pairs and the performance of quantum communications, but typically require collective operations, multiple entangled qubits or quantum memories,
making their near-term implementation highly demanding. 

In this work, we propose a method to recover the degree of entanglement distributed by entanglement swapping in noisy quantum repeater architectures. 
Our approach introduces a reversing operation based on weak measurement that probabilistically undoes the effect of decoherence on a single entangled pair, enabling heralded recovery without requiring collective operations or quantum memory. 
Reversal of weak measurement has been extensively studied, both in terms of its fundamental properties related to information contents~\cite{Jordan10,Cheong2012,Lim2014P,Lee2021Q,Hong2022,Lee2021R}, and for practical applications in suppressing decoherence~\cite{Korotkov10,Kim2012,Xiao2010,Lim2014,Lim2014a,Im2021}. 
Such reversing operations have also been experimentally demonstrated in a variety of physical platforms, including photonic systems~\cite{Kim2009,JCLee2011}, ion traps~\cite{Schindler13}, and superconducting circuits~\cite{Korotkov06,Katz08}.
Here, we apply reversing operations to the direct recovery of entanglement in two typical repeater models, i.e., two-way based on entanglement swapping and one-way  based on a relay protocol based on quantum teleportation~\cite{Muralidharan16,Lee2019}.
As representative examples of decoherence, we consider amplitude-damping, which frequently arises in stationary repeater systems such as superconducting circuits~\cite{Yan2022}, as well as photon loss during transmission~\cite{Pirandola2017,QiChao2016,Valivarthi2016}. 
Depending on the physical setting, different implementations of the reversing operation can be employed. 
For instance, in photonic platforms, photon loss can be effectively counteracted using noiseless linear amplification (NLA)~\cite{Ralph2009,Blandino2012,Chrzanowski2014,Xiang2010,Zavatta2011,Blandino2015,Zhao2017,Winnel2020}.

We show that the degree of entanglement can be substantially recovered even under strong noise, with a success probability determined by the noise strength. 
Notably, our method enables conditional recovery of entanglement even in parameter regimes where the distributed entanglement would otherwise vanish due to entanglement sudden death. 
We further identify the optimal reversing strength for each  repeater model, and show that maximal entanglement recovery can be achieved when the reversing strength slightly exceeds the corresponding noise level.
We note that our protocol requires neither quantum memory nor collective operations across multiple entangled copies. Moreover, its heralded nature makes it readily compatible with other entanglement recovery methods such as entanglement purification and distillation. 
We believe that our approach provides an efficient way for extending the range of quantum communication in currently available and near-term quantum communication systems.


\section{Entanglement recovery in a Single Entangled Pair}
\label{sec:SPmodel}

We begin by analyzing how amplitude-damping decoherence degrades the entanglement of an entangled pair and then introduce a direct method to probabilistically reverse the effect of noise. In realistic quantum communication channels, qubits inevitably undergo decoherence due to coupling with the environment. Among various types of noise, amplitude damping is particularly important  as it is the dominant decoherence mechanism in many stationary qubit platforms as well as flying qubits transmitted in optical channels. For example, amplitude damping manifests as photon loss in vacuum–single-photon qubits, spontaneous decay in atomic energy-level qubits, and zero-temperature energy relaxation in superconducting qubits.

\begin{figure}[t]
\includegraphics[width=0.8\linewidth]{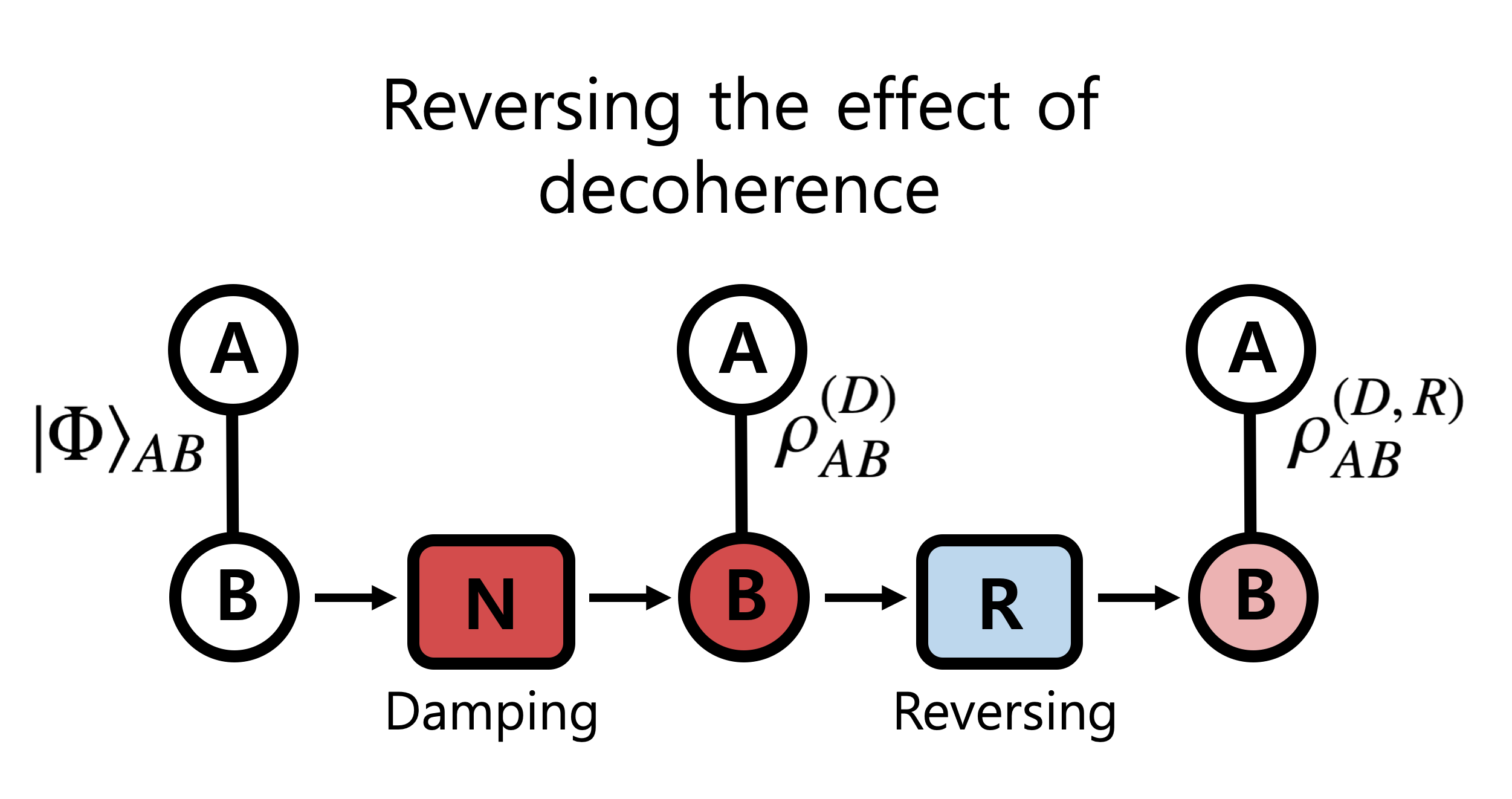}
\caption{Reversing the effect of decoherence on entangled pair.}
\label{fig:Schemesingp}
\end{figure}

Consider a qubit system ($S$) with basis states $\ket{0}_S$ and $\ket{1}_S$, and an environment ($E$) initially in $\ket{0}_E$. 
Amplitude damping decoherence, in which the excited state $\ket{1}$ is irreversibly and probabilistically transferred to the ground state $\ket{0}$ due to the state-dependent coupling between the system and the environment. The process can be modeled by the quantum map~\cite{Kim2012}:
\begin{align}
|0\rangle_S|0\rangle_E &\rightarrow |0\rangle_S|0\rangle_E, \nonumber\\
|1\rangle_S|0\rangle_E &\rightarrow \sqrt{1-D}\,|1\rangle_S|0\rangle_E + \sqrt{D}\,|0\rangle_S|1\rangle_E,
\label{eq:damping_map}
\end{align}
where $0 \le D \le 1$ denotes the damping strength determined by the system–environment coupling. 
Equivalently, this model can be represented as a quantum noise channel characterized by the Kraus operators
\begin{align}
\hat{D}_{1} &= \ket{0}\bra{0} + \sqrt{1-D}\ket{1}\bra{1}, \nonumber\\
\hat{D}_{2} &= \sqrt{D}\ket{0}\bra{1}.
\label{eq:Kraus_damping}
\end{align}
Here, $D=0$ corresponds to an ideal noiseless channel, while $D=1$ represents complete relaxation of the excited state $\ket{1}$.

Suppose that an arbitrary entangled pair $(A,B)$ is prepared in the state 
\begin{equation}
|\Phi\rangle_{AB} = \alpha|00\rangle_{AB} + \beta|11\rangle_{AB}, \qquad |\alpha|^2 + |\beta|^2 = 1.
\label{eq:bell_state1}
\end{equation}
We note that entangled states of the form $\alpha|01\rangle_{AB} + \beta|10\rangle_{AB}$ can be treated in a similar manner but are omitted here. 

If qubit $B$ undergoes amplitude damping decoherence as illustrated in Fig.~\ref{fig:Schemesingp}, the output density matrix of the pair $(A,B)$ becomes
\begin{equation}
\rho_{AB}^{(D)} =
\begin{pmatrix}
|\alpha|^2  & 0 & 0 & \alpha\beta^*\sqrt{1-D} \\
0 & 0 & 0 & 0 \\
0 & 0 & D|\beta|^2 & 0 \\
\alpha^*\beta\sqrt{1-D} & 0 & 0 & (1-D)|\beta|^2
\end{pmatrix}.
\label{eq:rho_damping1}
\end{equation}

To counteract the amplitude damping noise, we employ a reversing operation based on a weak measurement, a method that has been experimentally demonstrated in many previous works~\cite{Korotkov10,Kim2012,Xiao2010,Lim2014,Im2021,Kim2009,JCLee2011,Schindler13,Korotkov06,Katz08}. 
Unlike earlier approaches to suppress decoherence~\cite{Kim2012}, in which a weak measurement and its corresponding reversal operation are applied respectively before and after the qubits experience noise, our method applies a single reversing operation \emph{after} the entangled qubits have been distributed as illustrated in Fig.~\ref{fig:Schemesingp}.
This enables the recovery of degraded entanglement without requiring any pre-processing on the initially generated entangled pairs. 
The detailed procedure of the reversing protocol is described below.

To model the reversing operation, we consider a weak measurement that corresponds to the amplitude damping noise characterized by the Kraus operators in Eq.~(\ref{eq:Kraus_damping}). 
The reversing operation is implemented through the two Kraus operators
\begin{align}
\hat{R}_{1} &= \sqrt{1-R}\,\ket{0}\bra{0} + \ket{1}\bra{1}, \nonumber\\
\hat{R}_{2} &= \sqrt{R}\,\ket{1}\bra{0},
\label{eq:Kraus_reversing}
\end{align}
where $R$ denotes the strength of the reversing operation. 
These operators satisfy the completeness relation 
$\hat{R}_{1}^{\dagger}\hat{R}_{1} + \hat{R}_{2}^{\dagger}\hat{R}_{2} = I$, 
and correspond to two distinct measurement outcomes. 
A measurement outcome associated with $\hat{R}_{1}$ indicates a successful reversal, 
while the outcome associated with $\hat{R}_{2}$ corresponds to a failure event. 
Therefore, the reversing process is inherently probabilistic,
and its success probability as well as the amount of recovered entanglement depends on the reversing strength $R$.

Conditioned on the successful reversal, the density matrix of the pair $(A,B)$ becomes 
\begin{equation}
\rho_{AB}^{(D, R)}
=
\frac{1}{P_{AB}^{(D,R)}}
\begin{pmatrix}
\bar{R}|\alpha|^{2} & 0 & 0 & \alpha\beta^{*}\sqrt{\bar{D}\bar{R}} \\
0 & 0 & 0 & 0 \\
0 & 0 & D\bar{R}|\beta|^{2} & 0 \\
\alpha^{*}\beta\sqrt{\bar{D}\bar{R}} & 0 & 0 & \bar{D}|\beta|^{2}
\end{pmatrix},
\label{eq:rho_damping_rev1}
\end{equation}
where $\bar{R} =1-R$, $\bar{D}=1-D$.
The success probability of the reversing operation is given by
\begin{equation}
P_{AB}^{(D,R)}=\bar{R}|\alpha|^{2}+D\bar{R}|\beta|^{2}+\bar{D}|\beta|^{2}.
\label{eq:prob_damping_rev}
\end{equation}

\begin{figure}[t]
\centering
\raisebox{27ex}{\makebox[0pt][r]{\large\normalsize (a)\hspace{0.6em}}}%
\includegraphics[width=0.8\linewidth]{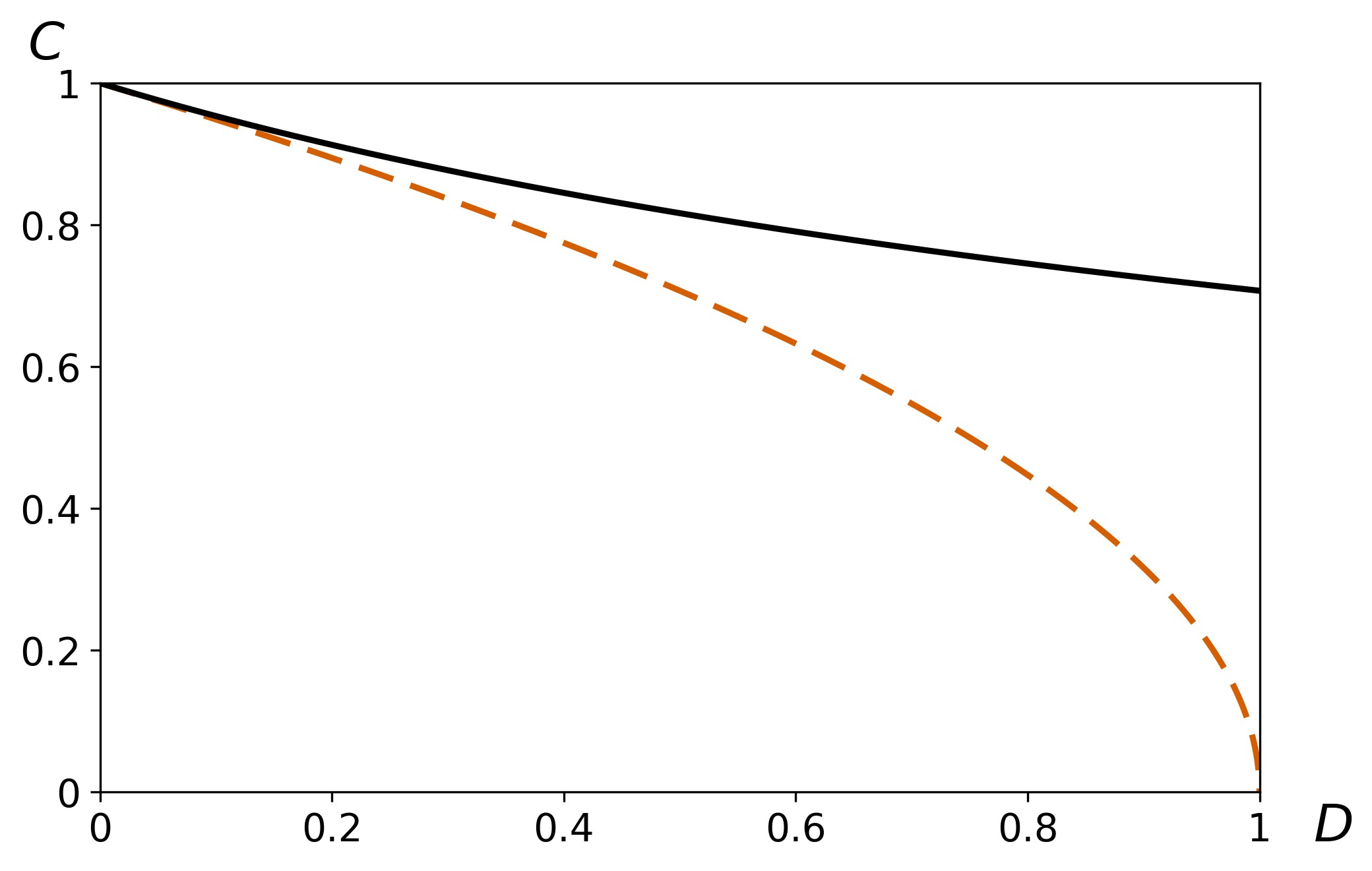}
\raisebox{27ex}{\makebox[0pt][r]{\large\normalsize (b)\hspace{0.6em}}}%
\includegraphics[width=0.8\linewidth]{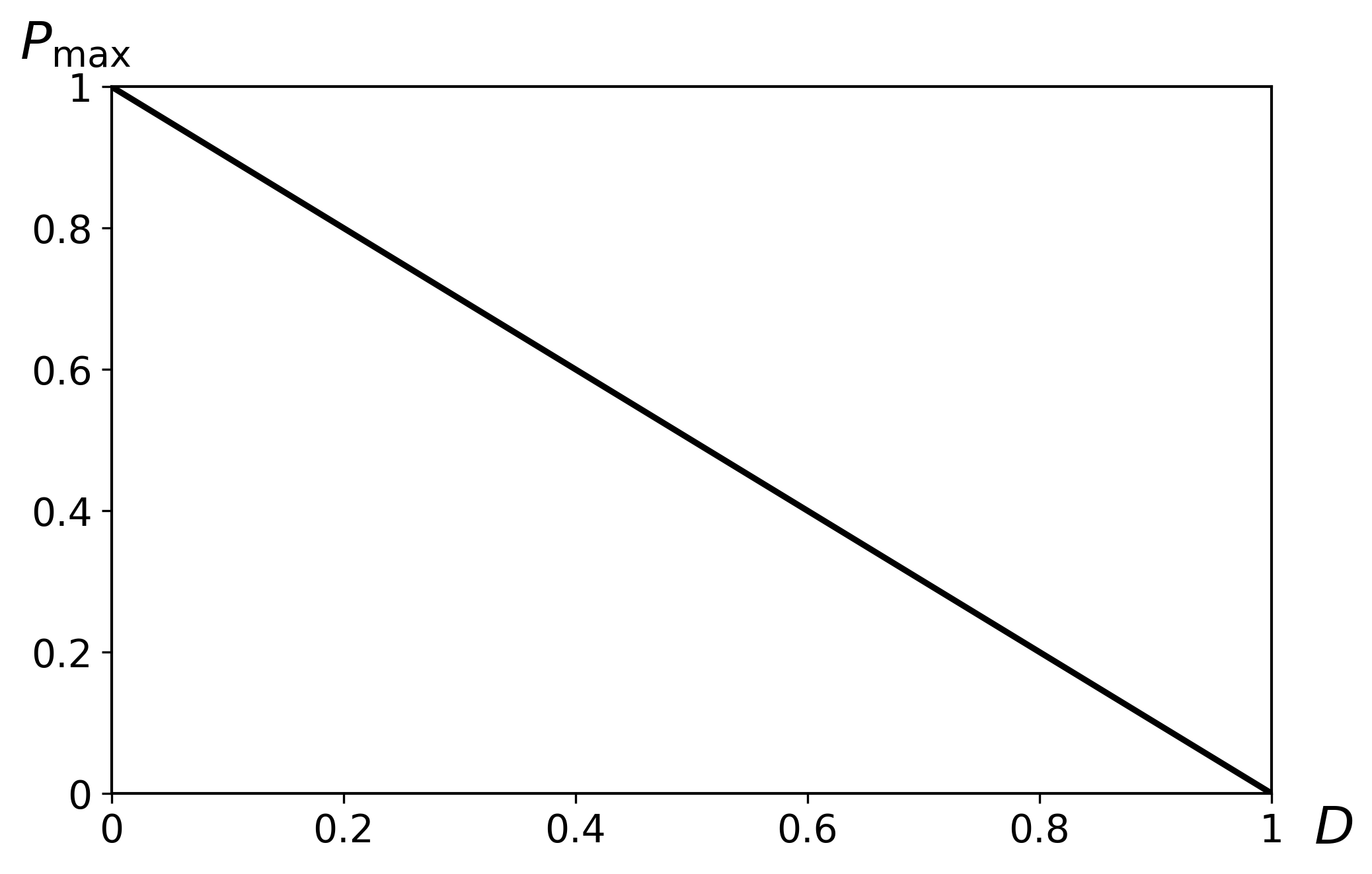}
\caption{(a) Concurrence of a single entangled pair under amplitude-damping decoherence. The dashed line shows the decay of concurrence as a function of the damping strength $D$, while the solid line shows the recovered concurrence obtained by applying the optimal reversing operation.
(b) Success probability of the optimal reversing operation as a function of $D$.
}
\label{fig:Concsingp}
\end{figure}

In order to quantify the degree of entanglement, we employ the concurrence~\cite{Wootters1998}, which serves as a convenient measure for bipartite entanglement in two-qubit systems. We can calculate the concurrence of the output state $\rho_{AB}^{(D,R)}$ as
\begin{equation}
C^{(D,R)}_{AB}=\frac{2|\alpha\beta|\sqrt{\bar{D}\bar{R}}}{P_{AB}^{(D,R)}},
\label{eq:con_sing_rev}
\end{equation} 
which for the maximally entangled initial pairs ($\alpha=\beta=1/\sqrt{2}$), i.e.,~the Bell pairs, becomes 
\begin{equation}
C^{(D,R)}_{AB}=\frac{2\sqrt{\bar{D}\bar{R}}}{2-R(1+D)}.
\end{equation}
Without applying the reversing operation, the concurrence is given by 
$C^{(D)}_{AB}=2|\alpha\beta|\sqrt{1-D}$. 
For the Bell pair input, this reduces to $C^{(D)}_{AB}=\sqrt{1-D}$, 
which monotonically decreases with increasing $D$, as shown in Fig.~\ref{fig:Concsingp}(a). 
When the reversing operation is applied, an appropriate choice of the reversing strength $R$ enables the recovered concurrence $C^{(D,R)}_{AB}$ to exceed the unrecovered value $C^{(D)}_{AB}$. 
The reversing strength that maximizes the concurrence can be obtained as
\begin{equation}
R_{\mathrm{opt}}^{(D)}=\frac{2D}{1+D},
\label{eq:optR}
\end{equation}
which yields the maximal recovered entanglement
\begin{equation}
C^{(D)}_{\mathrm{max}}=\frac{1}{\sqrt{1+D}}.
\end{equation}
The corresponding success probability can then be also obtained from Eq.~\eqref{eq:prob_damping_rev} and \eqref{eq:optR} as 
\begin{equation}
\label{eq:optProb}
P_{\mathrm{opt}}^{(D)}=1-\frac{1+D}{2}R_{\mathrm{opt}}^{(D)}=1-D.
\end{equation}
Notably, the optimal reversing strength obtained here in Eq.~\eqref{eq:optR} is always larger than the damping strength $R_{\mathrm{opt}}^{(D)} \geq D$ for all $0 \le D \le 1$. This reflects the fact that amplitude damping is an irreversible process that leaks coherence into the environment, whereas the reversing operation is probabilistic that conditionally amplifies the remaining coherence. More rigorously, after the application of the amplitude-damping channel described by the Kraus operators in Eq.~\eqref{eq:Kraus_damping}, not only the weight of $\ket{1}$ is attenuated through $\hat{D}_1$, but also $\ket{0}$ acquires additional population through $\hat{D}_2$. Consequently, a reversing strength $R$ equal to $D$ is insufficient, and a slightly stronger $R$ is required to compensate for this damping.

We plot the maximally recovered concurrence $C^{(D)}_{\mathrm{max}}$ by optimal reversing operation and the corresponding success probability by changing the damping strength in Fig.~\ref{fig:Concsingp}.
Importantly, the recovered entanglement always exceeds the unrecovered value since
$1/\sqrt{1+D} \ge \sqrt{1-D}$ for all $0 \le D \le 1$. 
This demonstrates that the reversing operation consistently enhances entanglement for any nonzero amount of damping, while its success probability decreases linearly and vanishes as $D \to 1$.

We can also employ the state fidelity to evaluate the performance of the reversing operation
\begin{equation}
F^{(D,R)}=\bra{\Phi^{+}}\rho_{AB}^{(D,R)}\ket{\Phi^{+}},
\end{equation}
which for a maximally entangled initial state becomes
\begin{equation}
F^{(D,R)}=
\frac{
\frac{1}{2}(2-R-D)+\sqrt{(1-D)(1-R)}
}{
2 - R(1+D)
}.
\end{equation}
The reversing strength maximizing the fidelity for a given $D$ can be thus obtained as
\begin{equation}
R_{\mathrm{f.opt}}^{(D)}=\frac{D(3+D)}{(1+D)^{2}},
\end{equation}
which is larger than the optimal reversing strength to maximize the concurrence, $R_{\mathrm{opt}}^{(D)}$ in Eq.~\eqref{eq:optR} for all $0<D<1$. 
This indicates that maximizing the fidelity requires a stronger reversing operation than maximizing the entanglement, reflecting the fact that full state restoration is a more stringent condition than recovering entanglement. In what follows, we therefore focus on entanglement recovery quantified by the concurrence, which is more suitable for the purposes of this work. We note that our analysis in this section shows that the reversing operation provides an useful tool to nontrivially recover entanglement that would otherwise be lost by amplitude damping.


\section{Entanglement recovery in a quantum repeater}
\label{sec:QRmodel}

In this section, we extend the reversing method to recover entanglement distributed through entanglement swapping in quantum repeater nodes. 
We consider two representative repeater models~\cite{Muralidharan16,Lee2019}.
The first is a two-way repeater, which extends the communication distance by performing entanglement swapping between two entangled qubit pairs. 
The second is a one-way repeater, which uses a relay protocol based on quantum teleportation to sequentially transfer one qubit of an entangled pair. 
The simplest configuration of each model consists of a single repeater node as illustrated in Fig.~\ref{fig:RepeaterModels}.

In the absence of noise, both models reduce to the same basic entanglement swapping procedure. 
We assume two maximally entangled Bell pairs: one between modes $A$ and $B$, $\ket{\Phi^{+}}_{AB} = (\ket{00} + \ket{11})/\sqrt{2}$, and another between modes $C$ and $D$, $\ket{\Phi^{+}}_{CD} = (\ket{00} + \ket{11})/\sqrt{2}$. 
A Bell-state measurement (BSM) on modes $B$ and $C$ projects the joint state onto the Bell basis, resulting in one of the Bell states shared between modes $A$ and $D$, conditioned on the BSM outcome (e.g., $\ket{\Phi^{+}}_{AD}$ for the outcome $\Phi^+$).  

In the following, we first examine how amplitude damping decoherence affects the distributed entanglement in a single-node repeater for both one-way and two-way protocols. 
We then apply the reversing operation to counteract the decoherence and recover the degraded entanglement.

\begin{figure}[t]
\includegraphics[width=1.0\linewidth]{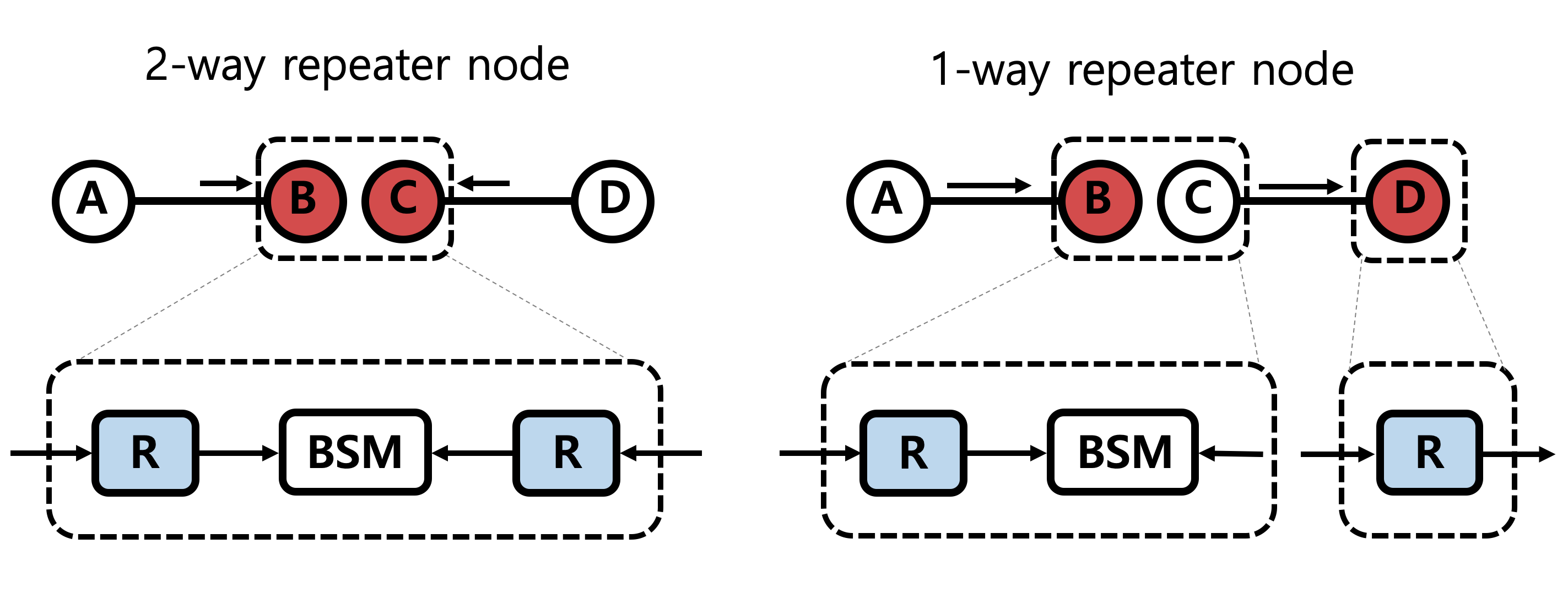}
\caption{Two quantum repeater models: a two-way protocol based on entanglement swapping and a one-way protocol based on quantum teleportation, each consisting of a single repeater node.
In the two-way model, the intermediate qubits $B$ and $C$ (shown in red) are assumed to undergo amplitude-damping decoherence before the Bell-state measurement (BSM).
In the one-way model, the transmitted qubits $B$ and $D$ in the relay teleportation protocol are assumed to experience amplitude damping (shown in red).
In our scheme, reversing operations are applied at each repeater node prior to the Bell-state measurements to undo the effects of noise.}
\label{fig:RepeaterModels}
\end{figure}

\subsection{Two-way repeater model}

\subsubsection{Decoherence effect}

In the two-way repeater model, which is based on entanglement swapping, we assume that the intermediate qubits $B$ and $C$ of two entangled pairs $(A,B)$ and $(C,D)$ undergo amplitude damping before BSM is performed on them, as illustrated in the left panel of Fig.~\ref{fig:RepeaterModels}.

We first consider an arbitrary entangled pair $(A,B)$ prepared in the state
\begin{equation}
|\Phi\rangle_{AB} = \alpha|00\rangle_{AB} + \beta|11\rangle_{AB}, \qquad |\alpha|^2 + |\beta|^2 = 1.
\label{eq:bell_state1}
\end{equation}
The amplitude damping acting on qubit $B$ modifies the density matrix of the pair $(A,B)$ into
\begin{equation}
\rho_{AB}^{(D_B)} =
\begin{pmatrix}
|\alpha|^2  & 0 & 0 & \alpha\beta^*\sqrt{1-D_B} \\
0 & 0 & 0 & 0 \\
0 & 0 & D_B|\beta|^2 & 0 \\
\alpha^*\beta\sqrt{1-D_B} & 0 & 0 & (1-D_B)|\beta|^2
\end{pmatrix},
\label{eq:rho_damping1}
\end{equation}
where $D_B$ denotes the damping strength acting on qubit $B$. 
Likewise, for the second entangled pair $(C,D)$ initially in
\begin{equation}
|\Phi\rangle_{CD} = \gamma|00\rangle_{CD} + \delta|11\rangle_{CD}, \qquad |\gamma|^2 + |\delta|^2 = 1,
\label{eq:bell_state2}
\end{equation}
the amplitude damping noise acting on qubit $C$ with $D_C$ yields
\begin{equation}
\rho_{CD}^{(D_C)} =
\begin{pmatrix}
|\gamma|^2  & 0 & 0 & \gamma\delta^*\sqrt{1-D_C} \\
0 & D_C|\delta|^2 & 0 & 0 \\
0 & 0 & 0 & 0 \\
\gamma^*\delta\sqrt{1-D_C} & 0 & 0 & (1-D_C)|\delta|^2
\end{pmatrix}.
\label{eq:rho_damping2}
\end{equation}

We consider the entangled state resulting from the entanglement-swapping operation. 
For simplicity, we assume that the damping strength acting on qubits $B$ and $C$ is identical, i.e., $D_B = D_C = D$. This assumption is reasonable when the same physical platform is used for both quantum repeaters.
After performing the BSM on qubits $B$ and $C$, the reduced density matrix of the distributed entangled pair $(A,D)$ for the outcomes 
$\Phi^{\pm}$ becomes
\begin{equation}
\rho_{\Phi}^{(D)} = 
\frac{1}{2B_{\Phi}^{(D)}}
\begin{pmatrix}
|\alpha\gamma|^2 & 0 & 0 & \pm \bar{D}\,\alpha\beta^{*}\gamma\delta^{*} \\
0 & D|\alpha\delta|^2 & 0 & 0 \\
0 & 0 & D|\beta\gamma|^2 & 0 \\
\pm \bar{D}\,\alpha^{*}\beta\gamma^{*}\delta & 0 & 0 & (\bar{D}^{2}+D^{2})|\beta\delta|^2
\end{pmatrix},
\label{eq:rho_AD_damping_phi}
\end{equation}
where $\bar{D}=1-D$ and 
\begin{equation}
B_{\Phi}^{(D)} = \tfrac12 \big( |\alpha\gamma|^{2} + D|\alpha\delta|^{2}
+ D|\beta\gamma|^{2} + (\bar{D}^{2}+D^{2})|\beta\delta|^{2} \big),
\end{equation}
is the probability of obtaining the outcome $\Phi^{\pm}$. Note that here the resulting states of $\Phi^{\pm}$ outcomes are associated with entanglement predominantly in the ${\ket{00},\ket{11}}$ subspace.

The concurrence of the resulting distributed state $\rho_{\Phi}^{(D)}$ in Eq.~\eqref{eq:rho_AD_damping_phi} is then calculated as
\begin{equation}
C_{\Phi}^{(D)}
= 
\max\!\left[
0,\,
\frac{ (1 - 2D)\,|\alpha\beta\gamma\delta| }{ B_{\Phi}^{(D)} }
\right].
\label{eq:concurrence_phi}
\end{equation}
For the maximally entangled initial pairs ($\alpha=\beta=\gamma=\delta=1/\sqrt{2}$), i.e., the Bell pairs, this reduces to
\begin{equation}
C_{\Phi}^{(D)} = \max\!\left[\,0,\, \frac{1 - 2D}{1 + D^{2}} \right].
\label{eq:conc_phi_simple}
\end{equation}
which we plot as the dashed line in Fig.~\ref{fig:Conc2way}(b) by varying $D$. 
We can observe that the concurrence decreases monotonically as $D$ increases. This clearly illustrates the degradation of distributed entanglement by amplitude damping decoherence.
In particular, the entanglement vanishes at $D=0.5$, indicating the occurrence of entanglement sudden death (ESD) by entanglement swapping.

For $\Psi^{\pm}$ outcomes of BSM, 
 the reduced density matrix is
\begin{equation}
\rho_{\Psi}^{(D)} = \frac{1}{2B_{\Psi}^{(D)}}
\begin{pmatrix}
0 & 0 & 0 & 0 \\
0 & \bar{D}|\alpha\delta|^2 & \pm \bar{D}\,\alpha\beta^{*}\gamma^{*}\delta & 0 \\
0 & \pm \bar{D}\,\alpha^{*}\beta\gamma\delta^{*} & \bar{D}|\beta\gamma|^2 & 0 \\
0 & 0 & 0 & 2\bar{D}D|\beta\delta|^2
\end{pmatrix},
\label{eq:rho_AD_damping_psi}
\end{equation}
with the corresponding probability
\begin{equation}
B_{\Psi}^{(D)} = \tfrac12 \big( \bar{D}|\alpha\delta|^{2}
+ \bar{D}|\beta\gamma|^{2}
+ 2\bar{D}D|\beta\delta|^{2} \big).
\label{eq:ProbBSM_psi_2way}
\end{equation}
This corresponds to entangled states predominantly in the subspace spanned by $\ket{01}$ and $\ket{10}$, i.e., states of the $\ket{01}\pm\ket{10}$ form.
Its concurrence can be obtained as
\begin{equation}
C_{\Psi}^{(D)}= \frac{ (1-D)\,|\alpha\beta\gamma\delta| }{ B_{\Psi}^{(D)} },
\label{eq:concurrence_psi}
\end{equation}
which reduces to $C_{\Psi}^{(D)} = (1+D)^{-1}$
for the maximally entangled Bell pair inputs. 
The concurrence decreases monotonically with increasing $D$, but remains positive for all $0 \leq D \leq 1$. Notably, the entanglement can already be enhanced at the level of a single entangled pair, whose concurrence is $\sqrt{1-D}$.
This indicates that the reduced states for the outcome $\Psi^{\pm}$ 
are more robust--in the sense that their entanglement never vanishes--than the reduced states for $\Phi^{\pm}$ against amplitude damping decoherence. 
However, we also note that the corresponding success probability $B_{\Psi}^{(D)}$ in Eq.~\eqref{eq:ProbBSM_psi_2way} decreases with increasing $D$ and vanishes in the limit $D \to 1$.

\subsubsection{Entanglement recovery by reversal}

In the two-way repeater model, the reversing operations are applied to the intermediate qubits $B$ and $C$, which have undergone amplitude damping prior to BSM as illustrated in the left panel of Fig.~\ref{fig:RepeaterModels}. 
When the reversing operation succeeds, the density matrix of the pair $(A,B)$ is transformed into 
\begin{equation}
\rho_{AB}^{(D_{B}, R_{B})}=\frac{1}{P_{AB}^{(D_{B},R_{B})}}
\begin{pmatrix}
\bar{R}_{B}|\alpha|^{2} & 0 & 0 & \alpha\beta^{*}\sqrt{\bar{D}_{B}\bar{R}_{B}} \\
0 & 0 & 0 & 0 \\
0 & 0 & D_{B}\bar{R}_{B}|\beta|^{2} & 0 \\
\alpha^{*}\beta\sqrt{\bar{D}_{B}\bar{R}_{B}} & 0 & 0 & \bar{D}_{B}|\beta|^{2}
\end{pmatrix},
\label{eq:rho_damping_rev1}
\end{equation}
where $R_{B}$ is the strength of the reversing operation applied to qubit $B$, and $\bar{R}_{B} = 1 - R_{B}$, $\bar{D}_{B}=1-D_{B}$. 
The success probability of the reversing operation is $P_{AB}^{(D_{B},R_{B})}=\bar{R}_{B}|\alpha|^{2}+D_{B}\bar{R}_{B}|\beta|^{2}+\bar{D}_{B}|\beta|^{2}$.
Similarly, conditioned on the success of the reversing operation on qubit $C$, the density matrix of the pair $(C,D)$ becomes
\begin{equation}
\rho_{CD}^{(D_{C}, R_{C})}=\frac{1}{P_{CD}^{(D_{C},R_{C})}}
\begin{pmatrix}
\bar{R}_{C}|\gamma|^{2} & 0 & 0 & \gamma\delta^{*}\sqrt{\bar{D}_{C}\bar{R}_{C}} \\
0 & D_{C}\bar{R}_{C}|\delta|^{2} & 0 & 0 \\
0 & 0 & 0 & 0 \\
\gamma^{*}\delta\sqrt{\bar{D}_{C}\bar{R}_{C}} & 0 & 0 & \bar{D}_{C}|\delta|^{2}
\end{pmatrix},
\label{eq:rho_damping_rev2}
\end{equation}
where $R_{C}$ is the strength of the reversing operation applied to qubit $C$ and
$P_{CD}^{(D_{C},R_{C})}=\bar{R}_{C}|\gamma|^{2}+D_{C}\bar{R}_{C}|\delta|^{2}+\bar{D}_{C}|\delta|^{2}$ is the corresponding success probability.

We now consider the resulting distributed state after the entanglement swapping. 
For simplicity, we assume that the damping strengths on qubits $B$ and $C$ are identical, 
$D_B = D_C = D$, and that the reversing operations applied to these qubits also have the same strength, 
$R_B = R_C = R$. 

Under these assumptions and conditioned on obtaining the BSM outcomes 
$\Phi^{\pm}$, the (unnormalized) density matrix of the distributed entangled pair $(A,D)$ becomes
\begin{equation}
\tilde{\rho}_{\Phi}^{(D,R)} =
\begin{pmatrix}
\bar{R}^{2}|\alpha\gamma|^{2} & 0 & 0 & 
\pm\, \bar{D}\bar{R}\,\alpha\beta^{*}\gamma\delta^{*} \\
0 & D\bar{R}^{2}|\alpha\delta|^{2} & 0 & 0 \\
0 & 0 & D\bar{R}^{2}|\beta\gamma|^{2} & 0 \\
\pm\, \bar{D}\bar{R}\,\alpha^{*}\beta\gamma^{*}\delta & 0 & 0 & 
(D^{2}\bar{R}^{2} + \bar{D}^{2})|\beta\delta|^{2}
\end{pmatrix}.
\label{eq:rho_AD_damping_rev_phi}
\end{equation}
The probability of obtaining the outcome $\Phi^{\pm}$ is
\begin{equation}
B_{\Phi}^{(D,R)}=
\frac{\bar{R}^{2}|\alpha\gamma|^{2}
+ D\bar{R}^{2}|\alpha\delta|^{2}
+ D\bar{R}^{2}|\beta\gamma|^{2}
+ (D^{2}\bar{R}^{2} + \bar{D}^{2})|\beta\delta|^{2}}{2 P_{AB}^{(D,R)} P_{CD}^{(D,R)}
}.
\label{eq:prob_AD_damping_rev_phi}
\end{equation}
We can then calculate the concurrence of the resulting state
\begin{equation}
C_{\Phi}^{(D,R)}=\max\!\left[0,\,
\frac{2\bar{D}\bar{R}|\alpha\beta\gamma\delta|
- 2D\bar{R}^{2}|\alpha\delta\,\beta\gamma|}{\mathrm{Tr}[\tilde{\rho}_{\Phi}^{(D,R)}]}
\right],
\label{eq:conc_phi_DR_general}
\end{equation}
which becomes 
\begin{equation}
C_{\Phi}^{(D,R)}=\max\!\left[0,\,
\frac{2\bar{R}(\bar{D}-D\bar{R})}{\bar{R}^2(1+D)^2+\bar{D}^2\,}
\right],
\label{eq:conc_maxinput_2way_phi}
\end{equation}
for the maximally entangled Bell pair inputs ($\alpha=\beta=\gamma=\delta=1/\sqrt{2}$).

\begin{figure*}[t]
\centering
\hspace{1.0em}
\begin{minipage}{0.32\textwidth}
\centering
\hspace{1.6em}
\begin{overpic}[width=0.98\linewidth]{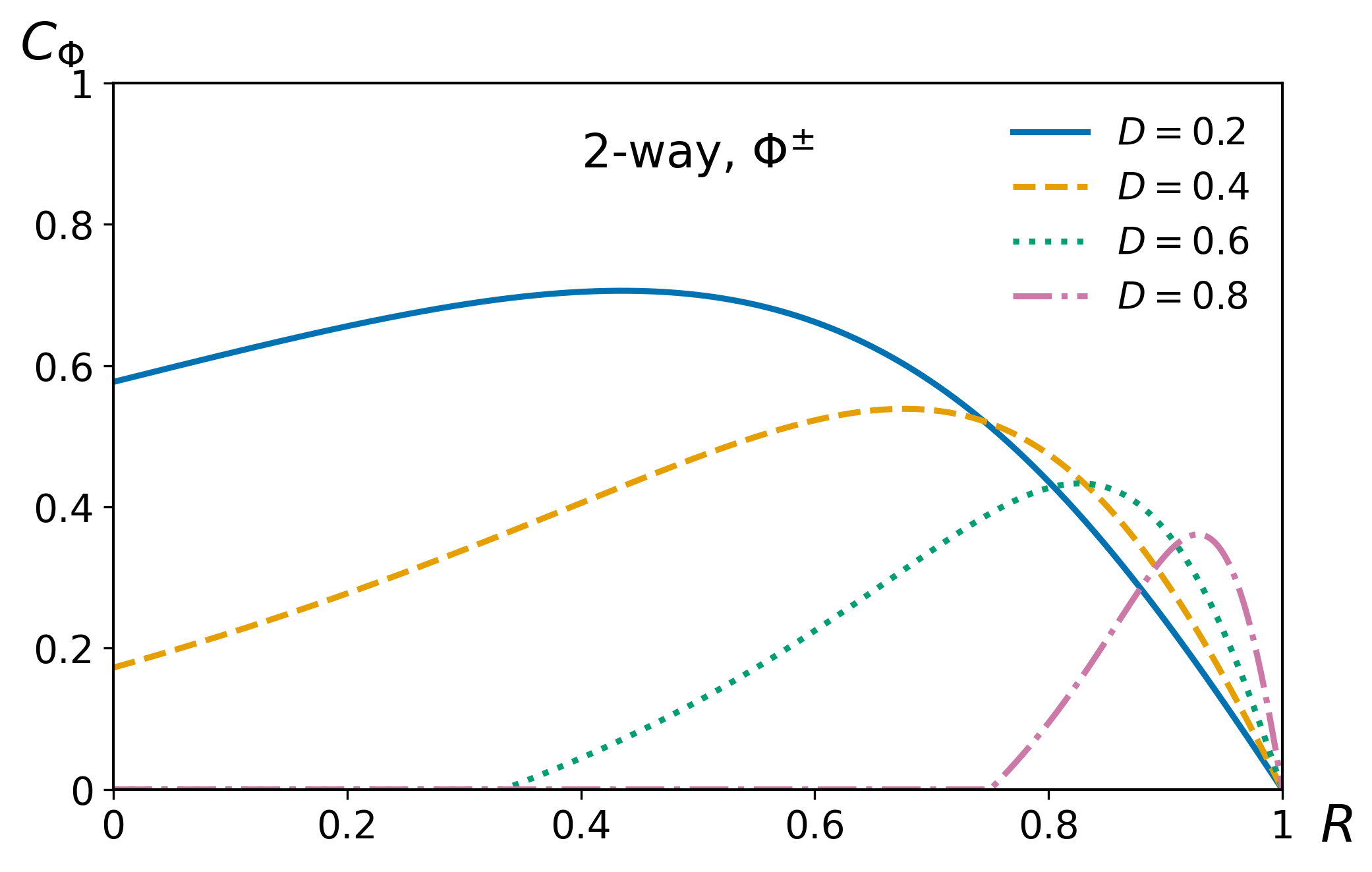}
  \put(-8,57){\normalsize (a)}
\end{overpic}
\end{minipage}\hfill
\begin{minipage}{0.32\textwidth}
\centering
\hspace{1.6em}
\begin{overpic}[width=0.98\linewidth]{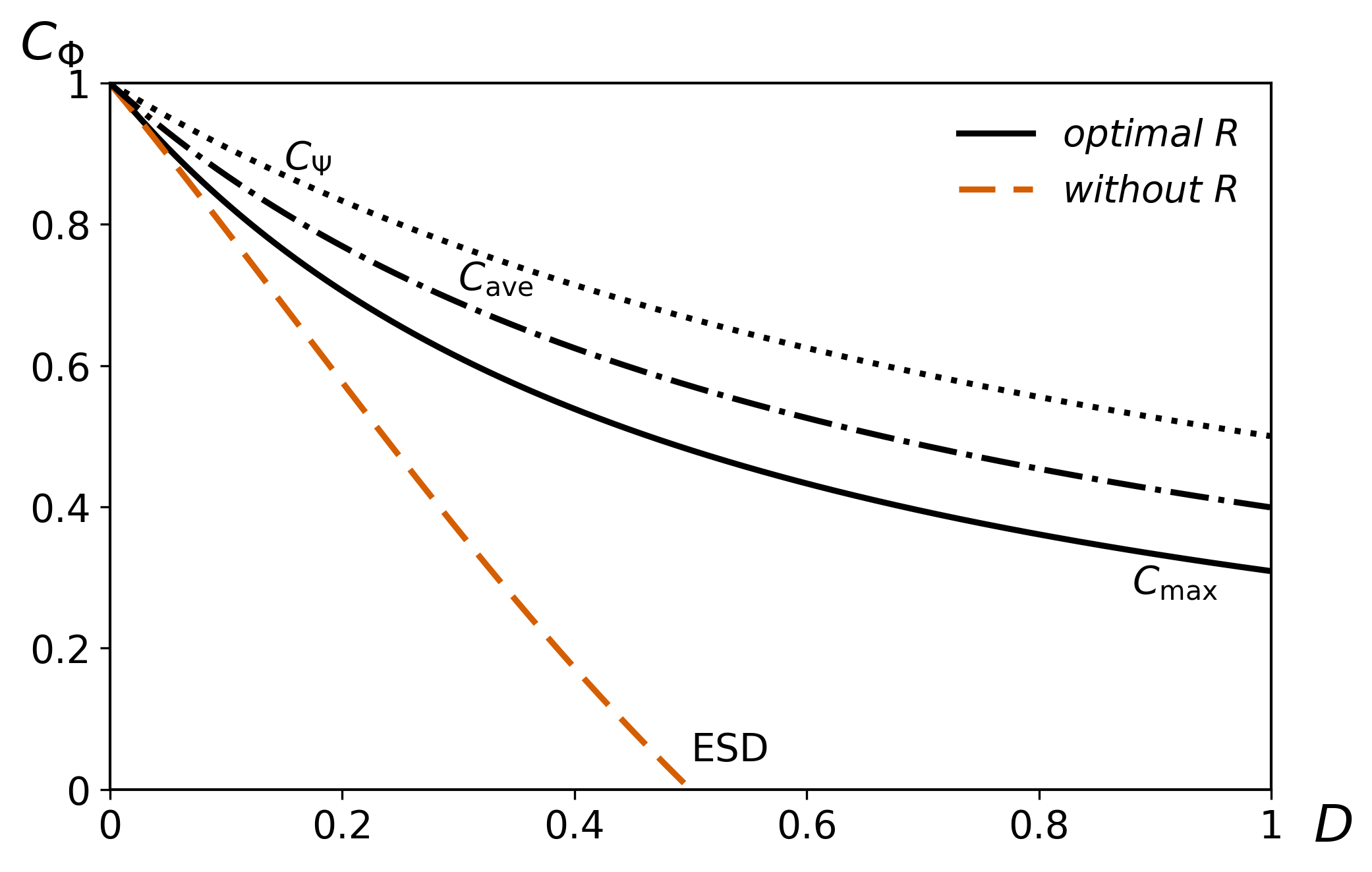}
  \put(-8,57){\normalsize (b)}
\end{overpic}
\end{minipage}\hfill
\begin{minipage}{0.32\textwidth}
\centering
\hspace{1.6em}
\begin{overpic}[width=0.98\linewidth]{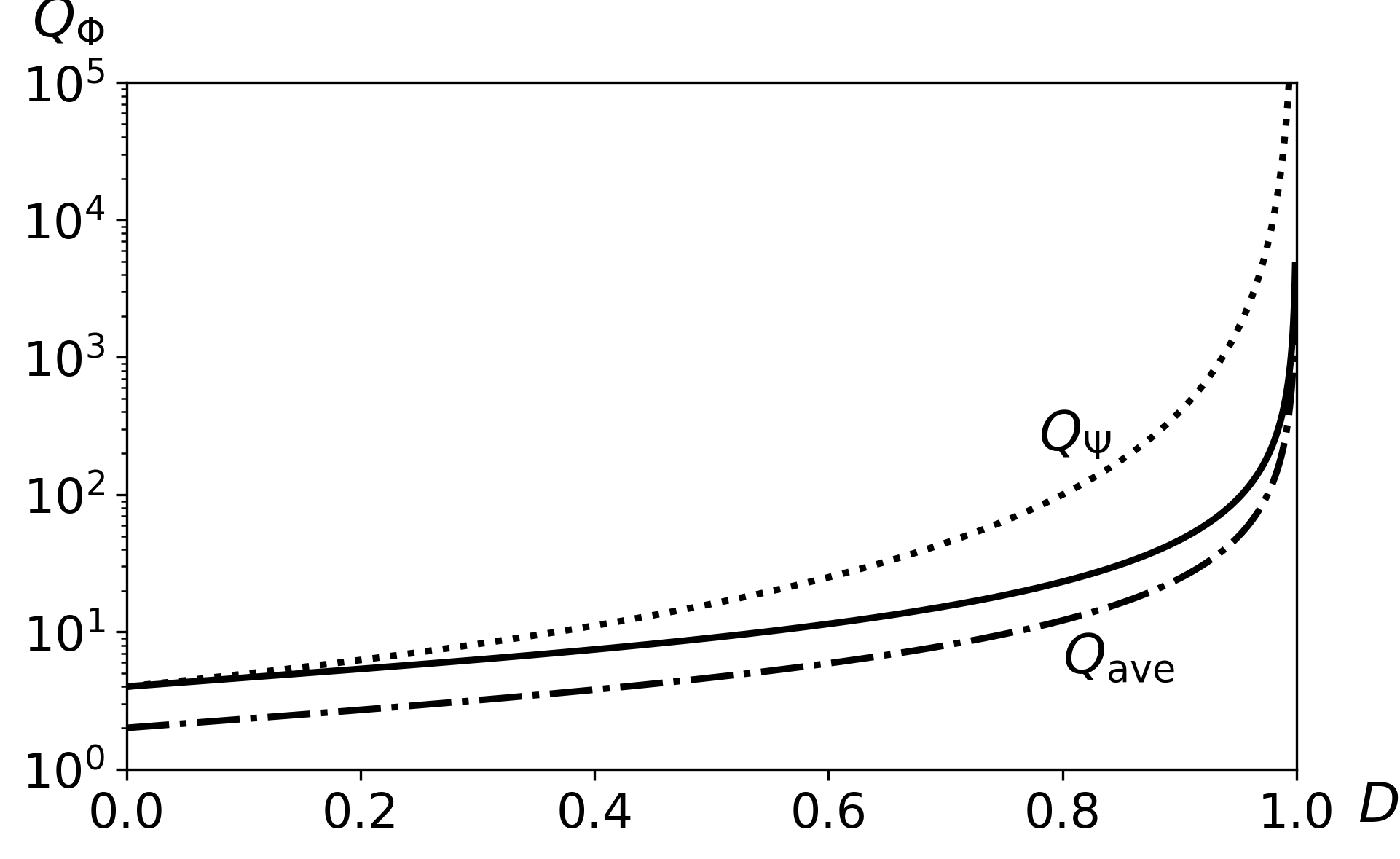}
  \put(-8,57){\normalsize (c)}
\end{overpic}
\end{minipage}

\caption{
(a)–(c) for the $\Phi^{\pm}$ outcomes in the two-way repeater model:
(a) Concurrence of the resulting distributed state as a function of $R$ for different damping strengths $D$.
(b) Concurrence by optimal reversing operation as a function of $D$, where the dashed line denotes the concurrence decay without reversing operation, which vanishes at $D=0.5$, while the solid line shows the recovered concurrence, demonstrating that entanglement can be effectively restored even in the damping region where entanglement sudden death (ESD) occurs. The concurrence for the $\Psi^{\pm}$ outcomes ($C_{\Psi}$) and the concurrence averaged over all outcomes ($C_{\mathrm{ave}}$) are shown as the dotted and dot-dashed lines, respectively, for comparison.
(c) Cost of Bell pairs required to recover entanglement via the optimal reversing operation for the $\Phi^{\pm}$ outcomes (solid line). The cost via entanglement swapping alone for the $\Psi^{\pm}$ outcomes ($Q_{\Psi}$) and the cost for obtaining the average concurrence over all outcomes ($Q_{\mathrm{ave}}$) are shown as the dotted and dot-dashed lines, respectively, for comparison.}
\label{fig:Conc2way}
\end{figure*}

Figure~\ref{fig:Conc2way}(a) shows the concurrence changes as a function of the reversing strength $R$ for several values of the damping strength $D$. 
We observe that the degraded entanglement, initially distributed from the maximally entangled inputs and subsequently affected by amplitude damping decoherence, can be substantially recovered by applying the reversing operation. 
Interestingly, the concurrence is maximized when $R$ is chosen slightly larger than $D$, as we discussed in Sec.~\ref{sec:SPmodel}. 
More precisely, the optimal reversing strength that maximizes the recovered concurrence can be calculated from Eq.~\eqref{eq:conc_maxinput_2way_phi} as
\begin{equation}
\bar{R}_{\Phi,\mathrm{opt}}^{(D)}
=\frac{1-D}{(1+D)^{2}}
\left[-D + \sqrt{1+2D+2D^{2}}\right].
\label{eq:Rbar_opt_phi}
\end{equation}
The corresponding maximally recovered concurrence is
\begin{equation}
C_{\Phi,\max}^{(D)}=
\frac{\sqrt{1+2D+2D^{2}}-D}{(1+D)^{2}}.
\label{eq:Cmax_2way_phi}
\end{equation}
Note that the optimal reversing strength for maximizing the distributed entanglement, $R_{\mathrm{opt}}^{(D)}$ in Eq.~\eqref{eq:Rbar_opt_phi} differs from the one for maximizing the single pair entanglement given in Eq.~\eqref{eq:optR}. In contrast to the single pair case, the optimal reversing operation in the two-way repeater model is required to effectively compensate for the cumulative damping effects combined by entanglement swapping.

In Fig.~\ref{fig:Conc2way}(b), we plot the maximum recovered concurrence $C_{\Phi,\max}^{(D)}$ in Eq.~\eqref{eq:Cmax_2way_phi} by varying $D$, compared to the unrecovered concurrence in Eq.~\eqref{eq:conc_phi_simple}. 
The result clearly shows that a substantial amount of entanglement can be restored even after it has been strongly degraded by amplitude damping during entanglement swapping. 
Remarkably, entanglement is effectively recovered even in the damping regions where ESD occurs, i.e.,~$D>0.5$. For example, our reversing method recovers a substantial amount of entanglement up to $C_{\max}\approx0.47$ with $R\approx0.77$ for the damping strength $D=0.52$ where entanglement swapping would otherwise completely destroy the entanglement.

We can estimate the resource cost used in the process of entanglement recovery in a single repeater node. More specifically, the number of consumed Bell pairs can be estimated based on the success probability of reversing operation and the probability of each BSM outcome. From Eq.~\eqref{eq:prob_damping_rev} and \eqref{eq:Rbar_opt_phi}, the success probability of the optimal reversing operation applied to each Bell pair is given by
\begin{equation}
P_{\Phi,\mathrm{opt}}^{(D)}=1-\frac{1+D}{2}R_{\Phi,\mathrm{opt}}^{(D)}=\frac{1-D}{2(1+D)}\left[1+ \sqrt{1+2D+2D^{2}}\right],
\end{equation}
implying that on average $1/P_{\Phi,\mathrm{opt}}^{(D)}$ Bell pairs are consumed to obtain one recovered entangled pair. Since two such recovered pairs are required in entanglement swapping, this leads to  $2/P_{\Phi,\mathrm{opt}}^{(D)}$. The corresponding probability of BSM outcomes $\Phi^{\pm}$ can be obtained from Eq.~\eqref{eq:prob_AD_damping_rev_phi} as
\begin{equation}
B_{\Phi,\mathrm{opt}}^{(D)}=\frac{1}{8\big(P_{\Phi,\mathrm{opt}}^{(D)}\big)^2}\left[(1+D)^2 \big(\bar{R}_{\Phi,\mathrm{opt}}^{(D)}\big)^2 +(1-D)^2\right].
\end{equation}
As the $\Phi^{-}$ outcome can be converted into $\Phi^{+}$ by only a local phase correction, the effective success probability is $2B_{\Phi,\mathrm{opt}}^{(D)}$.
Therefore, the overall cost of Bell pairs for optimal entanglement recovery can be estimated as
\begin{equation}
Q_{\Phi,\mathrm{opt}}^{(D)}=\frac{1}{P_{\Phi,\mathrm{opt}}^{(D)}B_{\Phi,\mathrm{opt}}^{(D)}},
\label{eq:cost2way_phi}
\end{equation}
which we plot in Fig.~\ref{fig:Conc2way}(c) by varying $D$.

The cost of Bell pairs required to maximally recover entanglement using the optimal reversing operation increases as the damping strength grows. Especially, as $D$ approaches unity, the success probability of the reversing operation vanishes so that the number of required Bell pairs diverges. This reflects the trade-off between the amount of entanglement recovery and the cost of Bell pairs under strong decoherence. Nevertheless, for a wide range of damping strengths $D$, the entanglement can be efficiently restored with a moderate Bell pair overhead, even in the ESD regime. 
For example, at $D=0.52$, the entanglement can be recovered up to $C_{\max}\approx 0.47$ by consuming approximately $10$ Bell pairs.

Conditioned on obtaining the BSM outcomes $\Psi^{\pm}$, the resulting (unnormalized) density matrix of the pair $(A,D)$ is obtained as
\begin{equation}
\tilde{\rho}_{\Psi}^{(D,R)} =
\begin{pmatrix}
0 & 0 & 0 & 0 \\
0 & \bar{D}\bar{R}|\alpha\delta|^{2} & 
\pm\, \bar{D}\bar{R}\,\alpha\beta^{*}\gamma^{*}\delta & 0 \\
0 & \pm\, \bar{D}\bar{R}\,\alpha^{*}\beta\gamma\delta^{*} & 
\bar{D}\bar{R}|\beta\gamma|^{2} & 0 \\
0 & 0 & 0 & 
2D\bar{D}\bar{R}|\beta\delta|^{2}
\end{pmatrix},
\label{eq:rho_AD_damping_rev_psi}
\end{equation}
and the probability of obtaining the outcomes $\Psi^{\pm}$ is
\begin{equation}
B_{\Psi}^{(D,R)}
=\frac{
\bar{D}\bar{R}|\alpha\delta|^{2}
+ \bar{D}\bar{R}|\beta\gamma|^{2}
+ 2D\bar{D}\bar{R}|\beta\delta|^{2}
}{2 P_{AB}^{(D,R)} P_{CD}^{(D,R)}}.
\label{eq:prob_AD_damping_rev_psi}
\end{equation}
The concurrence of the resulting state is obtained as
\begin{equation}
C_{\Psi}^{(D,R)}=\frac{|\alpha\beta\gamma\delta|}{
|\alpha\delta|^{2}+ |\beta\gamma|^{2}+ 2D|\beta\delta|^{2}}.
\label{eq:conc_psi_general}
\end{equation}
which becomes $C_{\Psi}^{(D)}=(1+D)^{-1}$
for the maximally entangled Bell pair inputs. 
We note that in this case the concurrence is not dependent on $R$ so that entanglement cannot be enhanced by the reversing operation. This arises because the projection onto $\ket{01}+\ket{10}$ removes the $\ket{11}$ component targeted by the reversing operations on qubits $B$ and $C$. Nevertheless, the concurrence for the $\Psi^{\pm}$ outcomes remains higher than that of the resulting states for $\Phi^{\pm}$ outcomes as shown in Fig.~\ref{fig:Conc2way}(b).

Therefore, when the goal is to retain the resulting entangled states for the $\Psi^{\pm}$ outcomes, the reversing operation is unnecessary prior to the BSM. The corresponding cost of Bell pairs is then given by
\begin{equation}
Q_{\Psi}^{(D)}=\frac{1}{B_{\Psi}^{(D)}},
\end{equation}
which is plotted as the dotted line in Fig.~\ref{fig:Conc2way}(c). 
Notably, despite the absence of a reversing operation, this cost increases more rapidly and remains consistently higher than the optimal recovery cost for the $\Phi^{\pm}$ outcomes given in Eq.~\eqref{eq:cost2way_phi} over the entire range of $D$.

As a result, the reversing operation can be effectively employed to recover entanglement against ESD induced by amplitude damping decoherence in the two-way repeater model, at the expense of a finite cost in terms of the initial Bell pairs $\ket{\Phi^{+}}$.
The optimal recovery strategy depends on both the target form of the entanglement and the available resource constraints.
If the objective is to maximize the extracted entanglement, one may retain only the resulting states from entanglement swapping with the $\Psi^{\pm}$ outcomes, as these always yield higher entanglement than the corresponding $\Phi^{\pm}$ outcomes. If, instead, the goal is to obtain the maximum attainable entanglement in the $\ket{00}+\ket{11}$ form consistently with the initial Bell pair $\ket{\Phi^{+}}$, the optimal strategy is to retain the resulting states associated with the $\Phi^{\pm}$ outcomes after applying the corresponding optimal reversing operation, while discarding the $\Psi^{\pm}$ outcomes.
Under finite resource constraints, the optimal strategy can be determined by comparing the costs shown in Fig.~\ref{fig:Conc2way}(c). Interestingly, although the $\Phi^{\pm}$-based strategy involves an additional heralded reversing operation prior to entanglement swapping, its overall cost remains lower than that of the $\Psi^{\pm}$-based strategy over the entire range of $D$.

Finally, if the objective is to maximize the average entanglement yield regardless of the entanglement form (i.e., $\ket{00}+\ket{11}$ or $\ket{01}+\ket{10}$), all BSM outcomes may be retained. In this case, the concurrence averaged over all resulting states is given by
\begin{equation}
C_{\mathrm{ave}}^{(D)}=2B_{\Phi,\mathrm{opt}}^{(D)}C_{\Phi,\max}^{(D)}+(1-2B_{\Phi,\mathrm{opt}}^{(D)})\frac{1}{1+D},
\label{eq:Cave_2way}
\end{equation}
while the corresponding Bell pair cost is simply determined by the inverse of the success probability of the optimal reversing operation,
\begin{equation}
Q_{\mathrm{ave}}^{(D)}=\frac{2}{P_{\Phi,\mathrm{opt}}^{(D)}}.
\end{equation}
These quantities are plotted as the dot-dashed lines in Fig.~\ref{fig:Conc2way}(b) and (c), respectively, for comparison with the other strategies. In this approach, all BSM outcomes are retained without postselection, thereby reducing the overall cost while still achieving substantial entanglement recovery. Since the BSM outcomes can be discriminated in each trial, our protocol can be applied selectively according to the type of the resulting entangled state, enabling flexible use for different applications.


\subsection{One-way repeater model}

\subsubsection{Decoherence effect}

In the one-way repeater model, which is based on a relay teleportation protocol, the qubits transmitted through the channel, $B$ and $D$, are assumed to experience damping during transmission, as illustrated in the right panel of Fig.~\ref{fig:Schemesingp}.
The density matrix of the pair $(A,B)$, where the amplitude-damping map acts on qubit $B$, is given in Eq.~(\ref{eq:rho_damping1}).
Likewise, the density matrix of the pair $(C,D)$, where the amplitude-damping map acts on qubit $D$, can be obtained in the same manner by replacing $\alpha \rightarrow \gamma$ and $\beta \rightarrow \delta$ in Eq.~(\ref{eq:rho_damping1}).

\begin{figure*}[t]
\centering
\hspace{1.0em}
\begin{minipage}{0.32\textwidth}
\centering
\hspace{1.6em}
\begin{overpic}[width=0.95\linewidth]{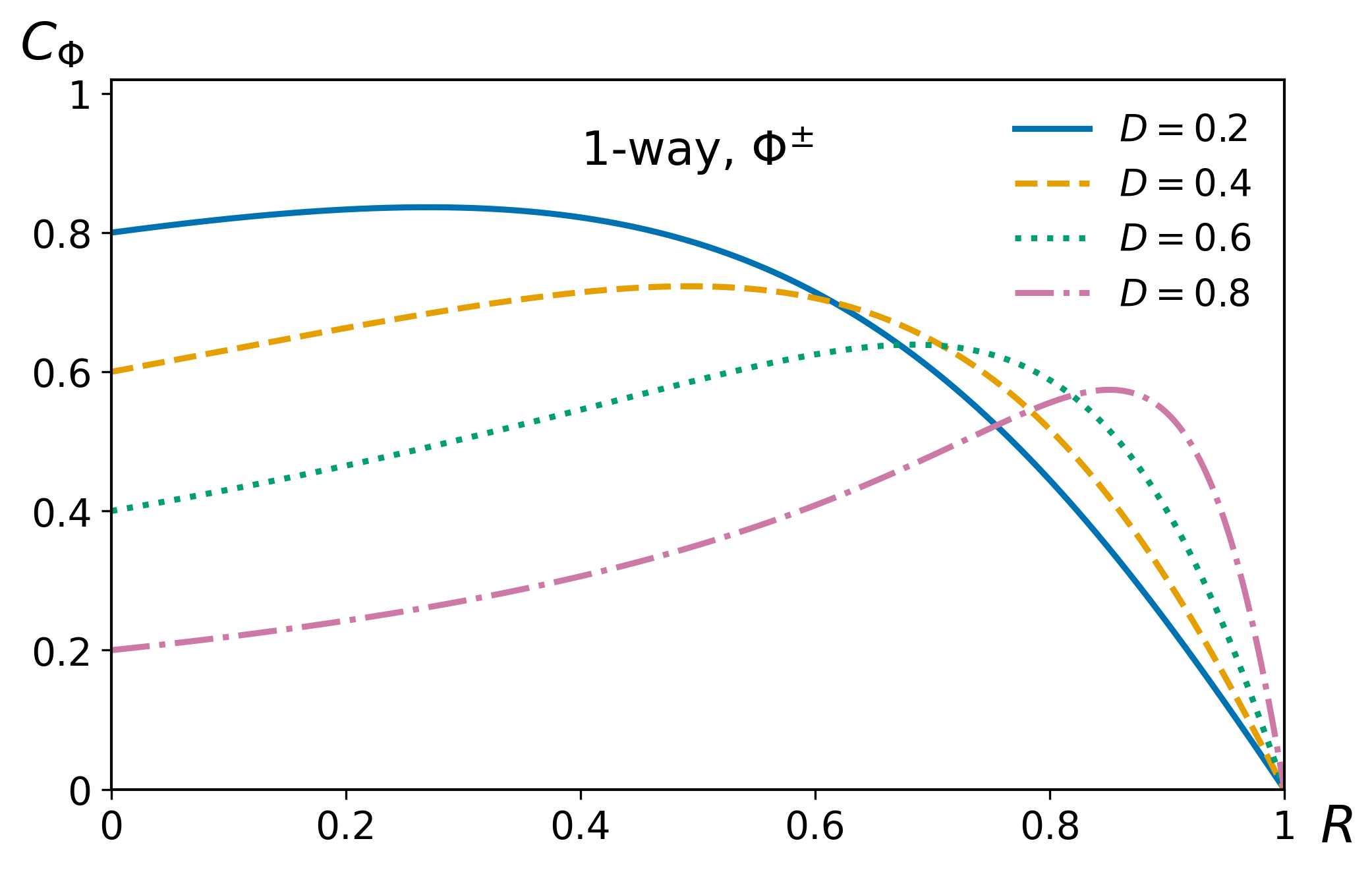}
  \put(-8,57){\normalsize (a)}
\end{overpic}
\end{minipage}\hfill
\begin{minipage}{0.32\textwidth}
\centering
\hspace{1.6em}
\begin{overpic}[width=0.95\linewidth]{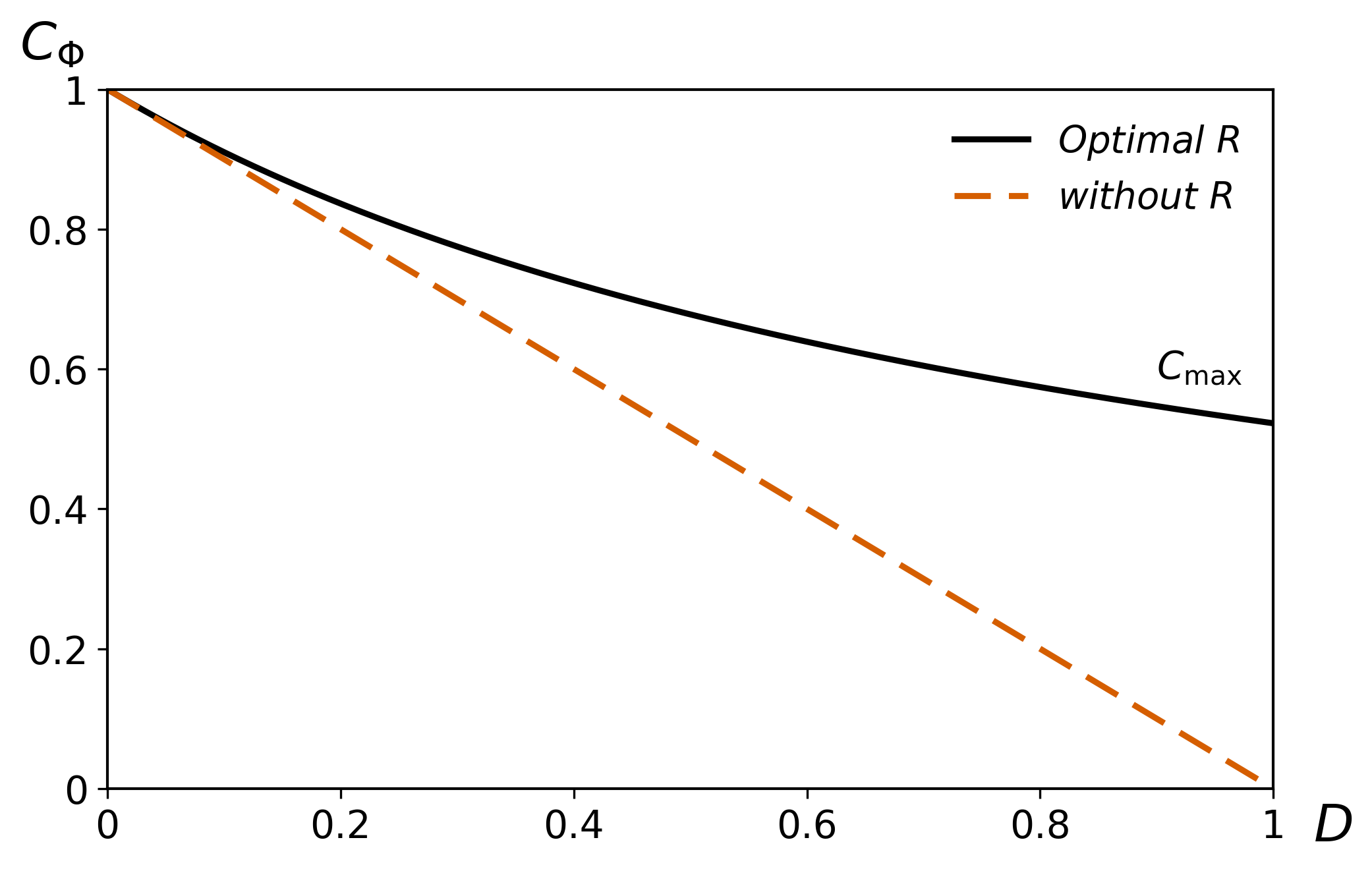}
  \put(-8,57){\normalsize (b)}
\end{overpic}
\end{minipage}\hfill
\begin{minipage}{0.32\textwidth}
\centering
\hspace{1.6em}
\begin{overpic}[width=0.95\linewidth]{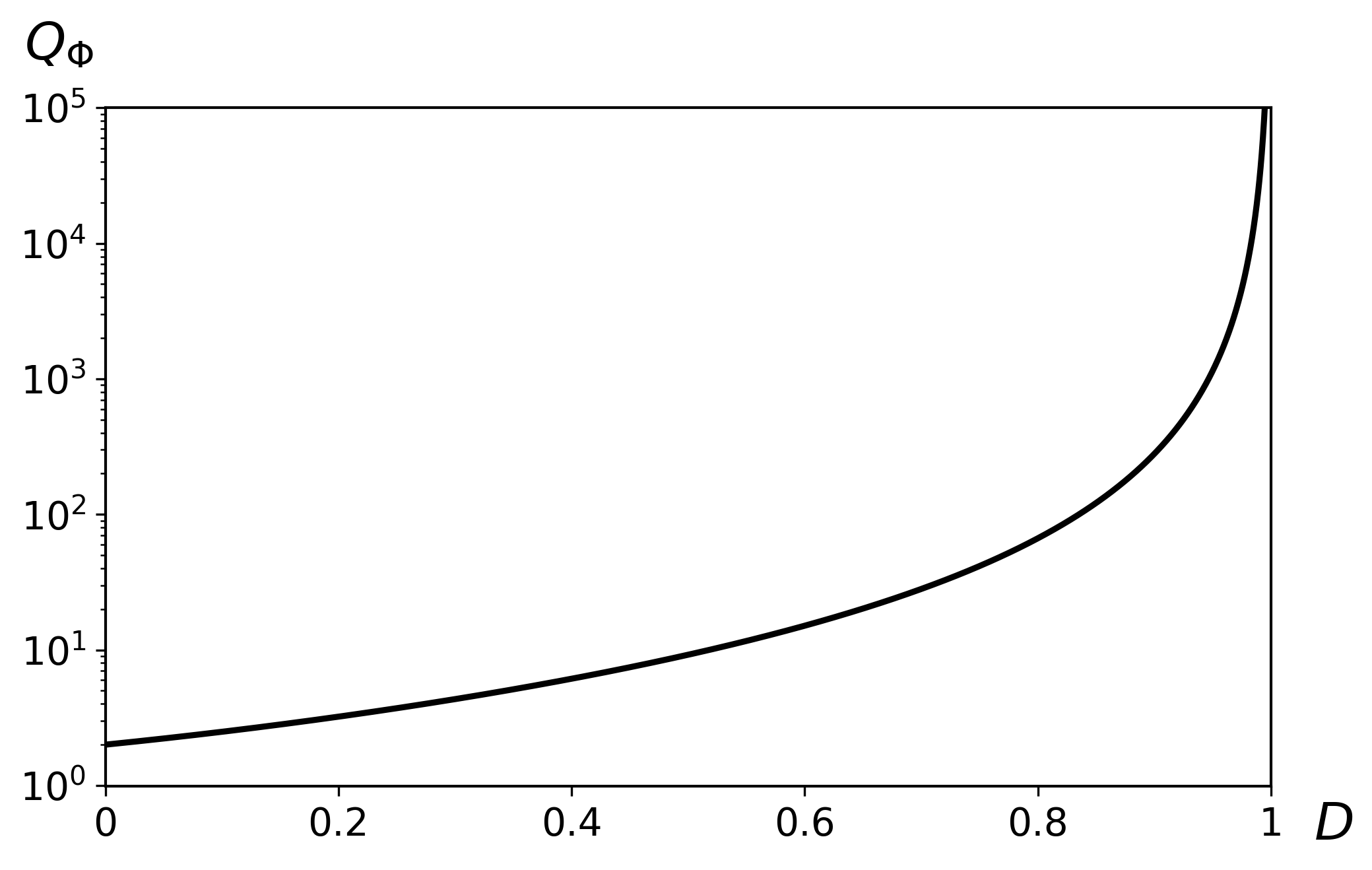}
  \put(-8,57){\normalsize (c)}
\end{overpic}
\end{minipage}

\vspace{1em}

\hspace{1.0em}
\begin{minipage}{0.32\textwidth}
\centering
\hspace{1.6em}
\begin{overpic}[width=0.95\linewidth]{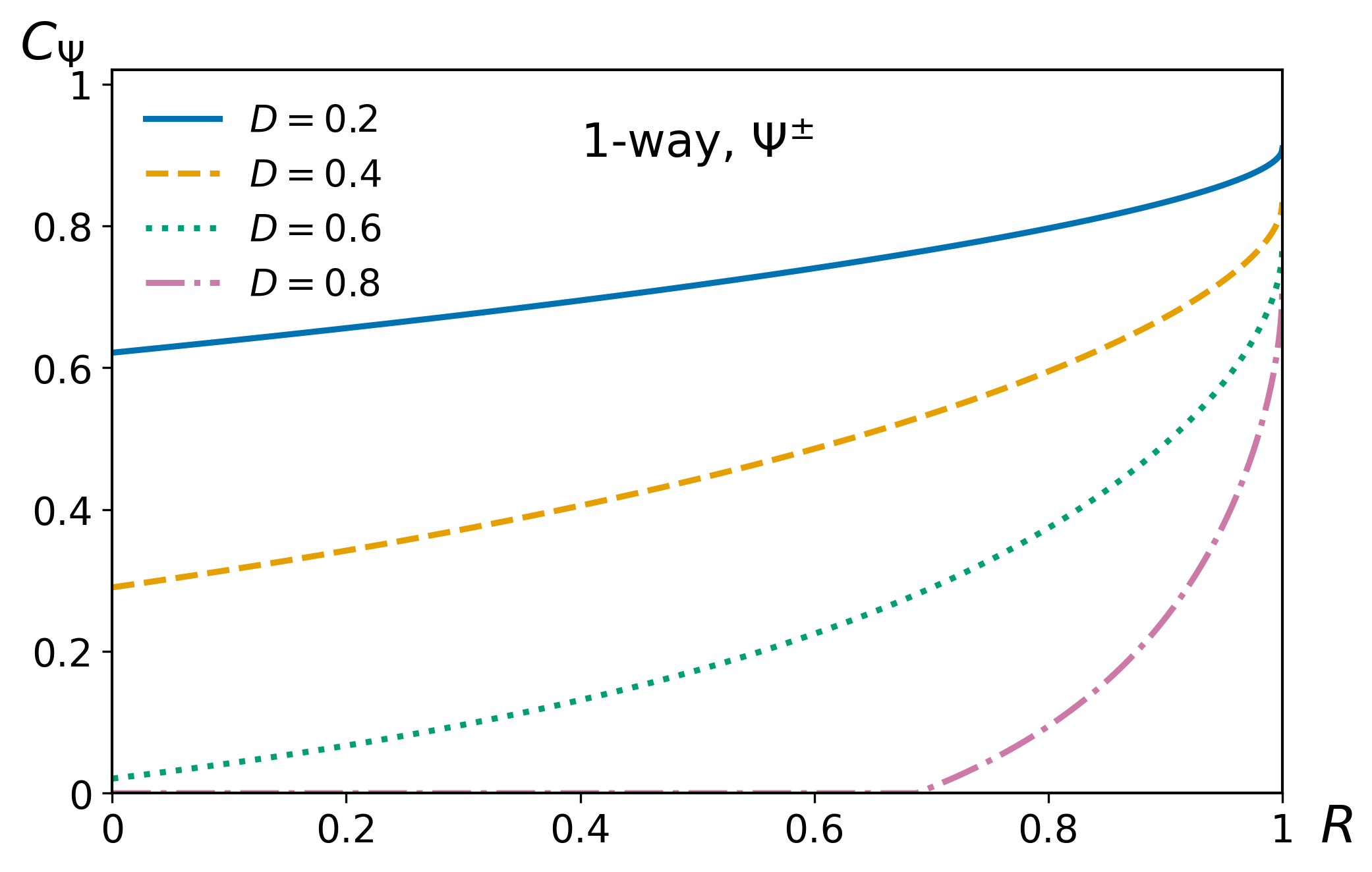}
  \put(-8,57){\normalsize (d)}
\end{overpic}
\end{minipage}\hfill
\begin{minipage}{0.32\textwidth}
\centering
\hspace{1.6em}
\begin{overpic}[width=0.95\linewidth]{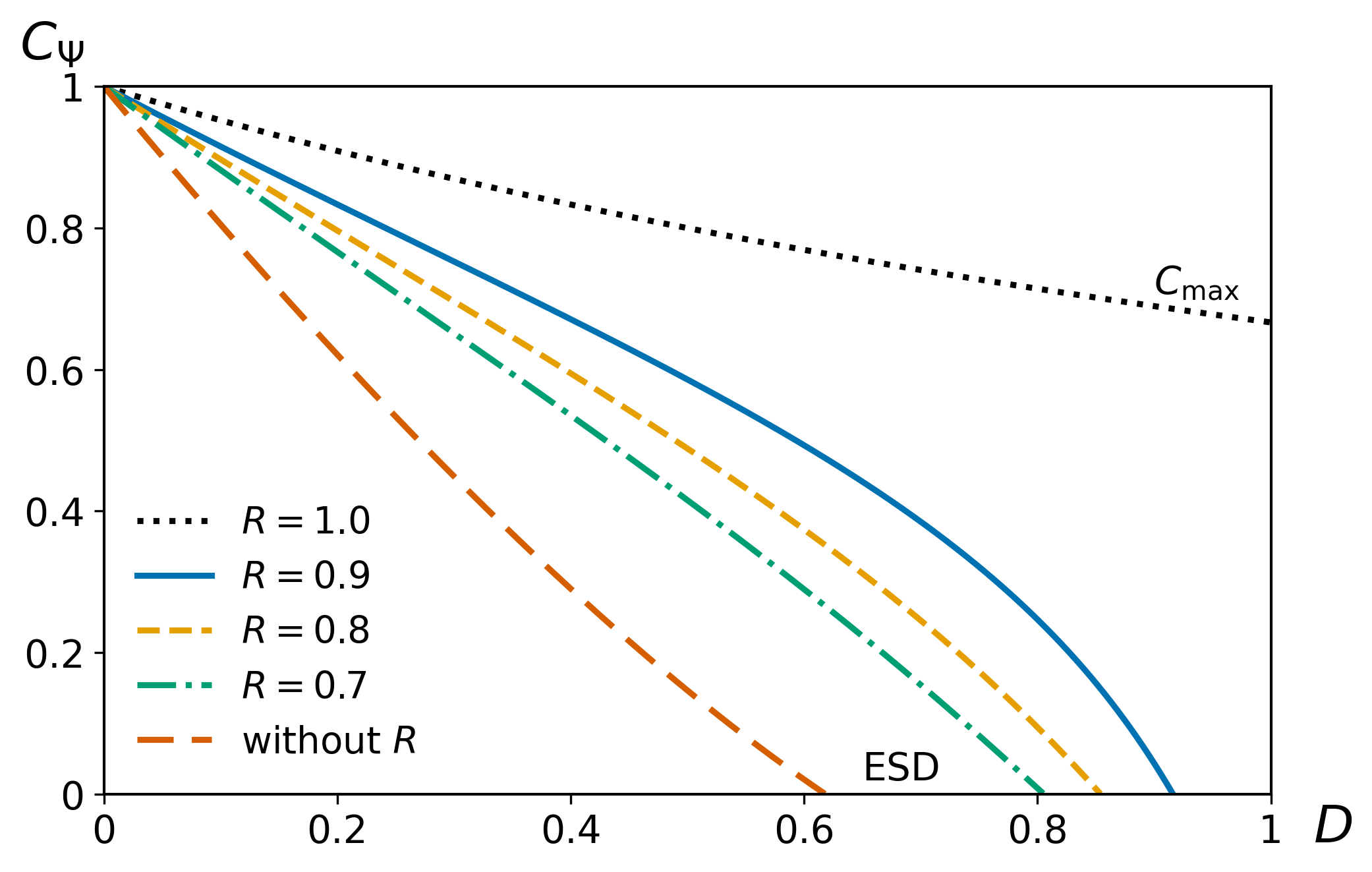}
  \put(-8,57){\normalsize (e)}
\end{overpic}
\end{minipage}\hfill
\begin{minipage}{0.32\textwidth}
\centering
\hspace{1.6em}
\begin{overpic}[width=0.95\linewidth]{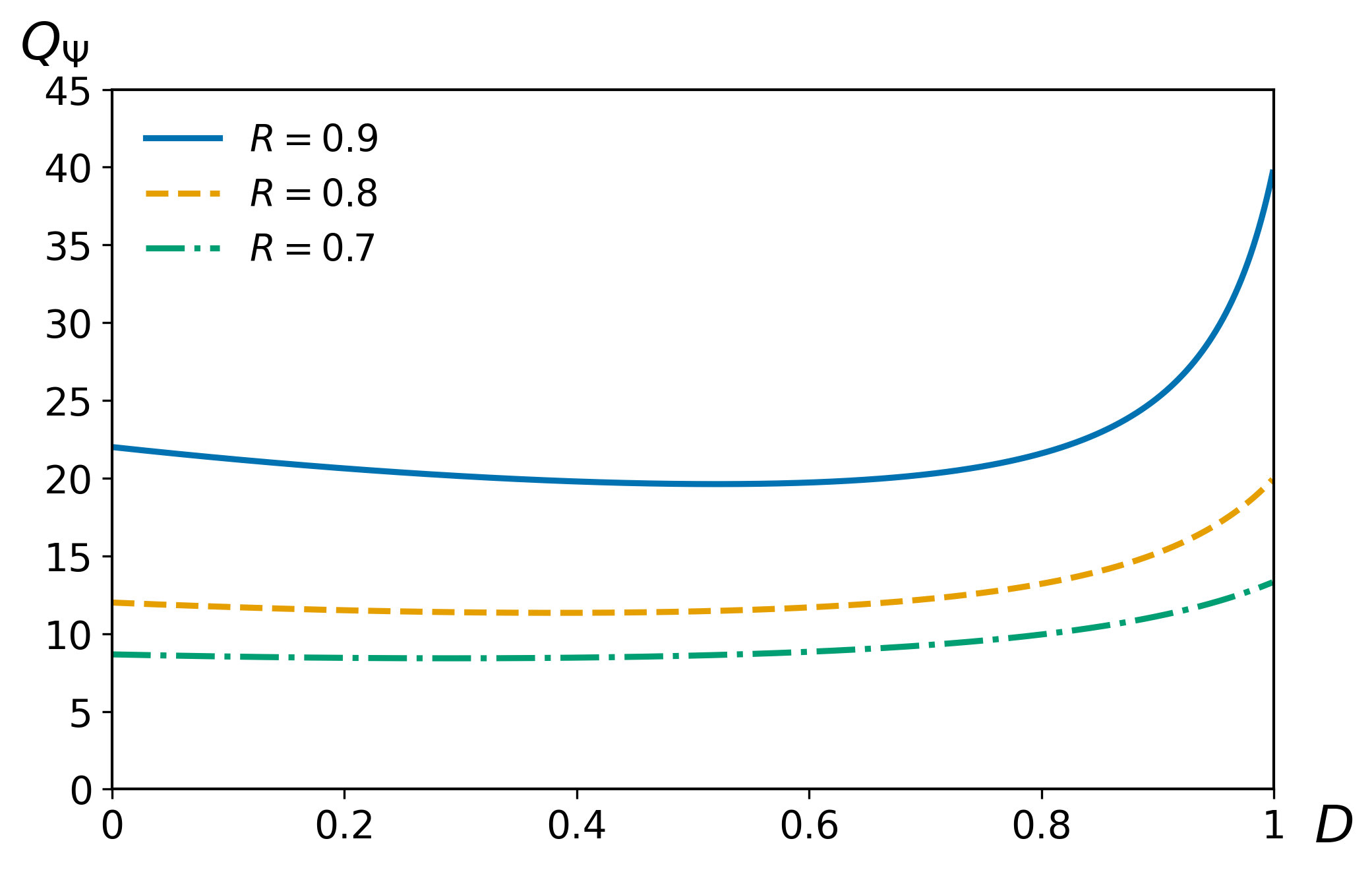}
  \put(-8,57){\normalsize (f)}
\end{overpic}
\end{minipage}

\caption{(a)–(c) for the $\Phi^{\pm}$ outcomes and (d)–(f) for the $\Psi^{\pm}$ outcomes in the one-way repeater model:
(a) Concurrence of the resulting distributed state as a function of $R$ for different damping strengths $D$.
(b) Concurrence by optimal reversing operation as a function of $D$, where the dashed and solid lines denote the unrecovered and recovered cases, respectively.
(c) Cost of Bell pairs required to recover entanglement by the optimal reversing operation against amplitude damping.
(d) Concurrence of the resulting distributed state as a function of $R$ for different damping strengths $D$.
(e) Recovered concurrence by reversing operation with different strengths $R$ as a function of the amplitude damping $D$, where the unrecovered concurrence vanishes at $D\approx0.618$.
(f) Cost of Bell pairs required to recover entanglement by the reversing operation with different strengths $R$ against amplitude damping.
}
\label{fig:Conc1way}
\end{figure*}

We now consider the resulting entangled state after the BSM is performed on qubits $B$ and $C$ in the relay teleportation protocol. 
We again assume that the amplitude damping strength acting on qubits $B$ and $D$ is identical, i.e., $D_B = D_D = D$. 

Under this assumption, the (reduced) density matrix of the distributed entangled pair $(A,D)$ for the outcomes $\Phi^{\pm}$ is
\begin{equation}
\rho_{\Phi}^{(D)} = 
\frac{1}{2B_{\Phi}^{(D)}}
\begin{pmatrix}
|\alpha\gamma|^2 & 0 & 0 & \pm \bar{D}\,\alpha\beta^{*}\gamma\delta^{*} \\
0 & 0 & 0 & 0 \\
0 & 0 & D|\beta\gamma|^2+\bar{D}D|\beta\delta|^2 & 0 \\
\pm \bar{D}\,\alpha^{*}\beta\gamma^{*}\delta & 0 & 0 & \bar{D}^{2}|\beta\delta|^2
\end{pmatrix},
\label{eq:rho_AD_damping_1w_phi}
\end{equation}
with the corresponding probability
\begin{equation}
B_{\Phi}^{(D)} 
= \tfrac12 \big( |\alpha\gamma|^{2} 
+ D|\beta\gamma|^{2} + \bar{D}|\beta\delta|^{2}\big).
\end{equation}
Its concurrence is given by
\begin{equation}
C_{\Phi}^{(D)} 
= \frac{ (1-D)|\alpha\beta\gamma\delta|}{B_{\Phi}^{(D)}},
\label{eq:conc_phi_1way_general}
\end{equation}
which becomes $C_{\Phi}^{(D)} = 1-D$ for maximally entangled Bell pair inputs. 

Likewise, for the outcomes $\Psi^{\pm}$, 
the reduced density matrix is obtainedㅊ as
\begin{equation}
\rho_{AD}^{(D)} = \frac{1}{2B_{\Psi}^{(D)}}
\begin{pmatrix}
D|\alpha\delta|^2 & 0 & 0 & 0 \\
0 & \bar{D}|\alpha\delta|^2 & \bar{D}\alpha\beta^*\gamma^*\delta & 0 \\
0 & \bar{D}\alpha^*\beta\gamma\delta^* & D^2|\beta\delta|^2 + \bar{D}|\beta\gamma|^2 & 0 \\
0 & 0 & 0 & \bar{D}D|\beta\delta|^2
\end{pmatrix},
\label{eq:rho_AD_damping_1w_psi}
\end{equation}
with the corresponding probability
\begin{equation}
B_{\Psi}^{(D)} 
= \tfrac12 \big(|\alpha\delta|^{2} 
+ \bar{D}|\beta\gamma|^{2} + D|\beta\delta|^{2} \big).
\end{equation}
Its concurrence is given by
\begin{equation}
C_{\Psi}^{(D)}=\max\bigg[0,\frac{\bar{D}|\alpha\beta\gamma\delta|
-D\sqrt{\bar{D}}\,|\alpha\beta||\delta|^{2}
}{B_{\Psi}^{(D)}}\bigg],
\label{eq:conc_psi_1way_general}
\end{equation}
which becomes $C_{\Psi}^{(D)} = \max\Big[
0, \,\bar{D}-D\sqrt{\bar{D}}\Big]$
for the maximally entangled Bell pair inputs. 

In Fig.~\ref{fig:Conc1way}(b) and Fig.~\ref{fig:Conc1way}(e), we plot the concurrence changes by increasing $D$ for both $\Phi^{\pm}$ and $\Psi^{\pm}$ outcomes (dashed lines) when the initial inputs are Bell pairs. Both represent a linear decrease of the entanglement as the amplitude damping strength $D$ increases, but the entanglement for $\Phi^{\pm}$ result remains positive for all $0 \leq D \leq 1$, while the entanglement for $\Psi^{\pm}$ result decreases more rapidly with $D$ than $\Phi^{\pm}$ and vanishes at 
$D_{\mathrm{ESD}} \simeq 0.618$. This reveals the difference of the robustness of the two BSM outcomes in one-way repeater model against amplitude damping decoherence.
In particular, unlike the two-way repeater model, the $\Phi^{\pm}$ outcomes are more robust than $\Psi^{\pm}$ over all range of damping strengths $D$.
This reveals the difference in the robustness of the two BSM outcomes in one-way repeater model against amplitude damping decoherence.
In particular, in the sense that the concurrence remains nonzero and no ESD occurs over the all range of $D$, the $\Phi^{\pm}$ outcomes are more robust than $\Psi^{\pm}$ in contrast to two-way repeater model, where the $\Psi^{\pm}$ outcomes exhibit higher robustness than $\Phi^{\pm}$.

\subsubsection{Entanglement recovery by reversal}

In the one-way repeater model, the reversing operations are applied to the qubits $B$ and $D$, which have transmitted through the channel and experienced damping as illustrated in the right panel of Fig.~\ref{fig:RepeaterModels}. 
The density matrix of the pair $(A,B)$ conditioned on the success of the reversing operation is given in Eq.~\eqref{eq:rho_damping_rev1}. Likewise, the density matrix of the pair $(C,D)$ after the reversing operation applied on $D$ can be obtained as the same form by replacing $\alpha \rightarrow \gamma$ and $\beta \rightarrow \delta$. We assume again that the damping strengths on qubits $B$ and $D$ are identical, $D_B = D_D = D$, as well as the reversing strengths $R_B = R_D = R$. 

Let us then consider the resulting state after the BSM is performed on qubits $B$ and $C$. 
Conditioned on obtaining the BSM outcomes 
$\Phi^{\pm}$, the (unnormalized) density matrix of the distributed entangled pair $(A,D)$ becomes
\begin{equation}
\tilde{\rho}_{\Phi}^{(D,R)} =
\begin{pmatrix}
\bar{R}^{2}|\alpha\gamma|^{2} & 0 & 0 & 
\pm\, \bar{D}\bar{R}\,\alpha\beta^{*}\gamma\delta^{*}\\
0 & 0 & 0 & 0 \\
0 & 0 & D\bar{R}^{2}|\beta\gamma|^{2} +D\bar{D}\bar{R}|\beta\delta|^{2}& 0 \\
\pm\, \bar{D}\bar{R}\,\alpha^{*}\beta\gamma^{*}\delta & 0 & 0 & 
\bar{D}^{2}|\beta\delta|^{2}
\end{pmatrix}.
\label{eq:rho_AD_damping_rev_phi_1way}
\end{equation}
The probability of obtaining the outcome $\Phi^{\pm}$ in BSM is
\begin{equation}
B_{\Phi}^{(D,R)}=
\frac{\bar{R}^{2}|\alpha\gamma|^{2}
+ D\bar{R}^{2}|\beta\gamma|^{2}
+ (D\bar{D}\bar{R} + \bar{D}^{2})|\beta\delta|^{2}}{2 P_{AB}^{(D,R)} P_{CD}^{(D,R)}
}.
\label{eq:prob_AD_damping_rev_phi_1way}
\end{equation}
The concurrence of the state in Eq.~\eqref{eq:rho_AD_damping_rev_phi_1way} can be calculated as
\begin{equation}
C_{\Phi}^{(D,R)}
=\frac{2\bar D \bar R\,|\alpha\beta\gamma\delta|}
     {\mathrm{Tr}[\tilde{\rho}_{\Phi}^{(D,R)}]},
\label{eq:conc_1way_phi_DR_general}
\end{equation}
which reduces to 
\begin{equation}
C_{\Phi}^{(D,R)}=
\frac{2\bar D \bar R}
{\bar R^{2}(1+D) + D\bar D\,\bar R + \bar D^{2}},
\label{eq:conc_1way_phi_maxent_reversing}
\end{equation}
for the maximally entangled Bell pair inputs.
In Fig.~\ref{fig:Conc1way}(a), we plot the recovered concurrence by changing $R$ for various $D$. We can observe the recovery of the concurrence with a peak at a certain $R$ for a given $D$. We can find analytically the optimal reversing strength that maximizes the recovered concurrence as
\begin{equation}
\bar R_{\Phi,\mathrm{opt}}^{(D)}=\frac{1-D}{\sqrt{1+D}},
\label{eq:Ropt_1w_phi}
\end{equation}
leading to the maximal recovered concurrence
\begin{equation}
C_{\Phi,\max}^{(D)}
=\frac{2}{D+2\sqrt{1+D}}.
\label{eq:Cmax_1w_phi}
\end{equation}
In Fig.~\ref{fig:Conc1way}(b), the maximum recovered concurrence is compared with the concurrence of the unrecovered entanglement by changing the damping strength $D$.

We can also estimate the number of consumed initial Bell pairs to achieve this maximum concurrence in a single one-way repeater node (by following the same procedure as in the two-way repeater model)
as
\begin{equation}
Q_{\Phi,\mathrm{opt}}^{(D)}=\frac{1}{P_{\Phi,\mathrm{opt}}^{(D)}B_{\Phi,\mathrm{opt}}^{(D)}},
\end{equation}
where the success probability of the optimal reversing operation is obtained from Eq.~\eqref{eq:prob_damping_rev} and \eqref{eq:Ropt_1w_phi} as
\begin{equation}
P_{\Phi,\mathrm{opt}}^{(D)}=1-\frac{1+D}{2}R_{\Phi,\mathrm{opt}}^{(D)}=\frac{\bar{D}}{2}\left[1+ \sqrt{1+D}\right],
\end{equation}
and 
\begin{equation}
B_{\Phi,\mathrm{opt}}^{(D)}=\frac{1}{8\big(P_{\Phi,\mathrm{opt}}^{(D)}\big)^2}\left[(1+D)\big(\bar{R}_{\Phi,\mathrm{opt}}^{(D)}\big)^2 +D\bar{D}\bar{R}_{\Phi,\mathrm{opt}}^{(D)}+\bar{D}^2\right]
\end{equation}
is the corresponding probability of BSM outcome $\Phi^+$ or $\Phi^-$. 
We plot the cost of Bell pair in Fig.~\ref{fig:Conc1way}(c), which exhibits the increase of cost as the damping strength $D$ grows.

\begin{table}
\centering

\caption{Summary of entanglement sudden death (ESD) and the effect of reversing operations in one-way and two-way repeater models under amplitude damping when the initial Bell pair is $\ket{\Phi^+}$.}
\label{tab:summary_esd_reversing}
\renewcommand{\arraystretch}{1.25}
\setlength{\tabcolsep}{8pt}
{
\begin{tabular}{lccc}
\hline\hline
Model & BSM outcome & ESD occurs? & Recovery by $R$ \\
\hline
\multirow{2}{*}{One-way} 
& $\Phi^{\pm}$ & No  & Yes \\
& $\Psi^{\pm}$ & Yes & Yes \\
\hline
\multirow{2}{*}{Two-way} 
& $\Phi^{\pm}$ & Yes & Yes \\
& $\Psi^{\pm}$ & No  & No \\
\hline\hline
\end{tabular}
}
\end{table}

For the BSM outcomes $\Psi^{\pm}$, the resulting (unnormalized) density matrix of the pair $(A,D)$ is obtained as
\begin{equation}
\tilde{\rho}_{\Psi}^{(D,R)} =
\begin{pmatrix}
D\bar{R}^2|\alpha\delta|^{2} & 0 & 0 & 0 \\
0 & \bar{D}\bar{R}|\alpha\delta|^{2} & 
\pm\bar{D}\bar{R}\alpha\beta^{*}\gamma^{*}\delta & 0 \\
0 & \pm\bar{D}\bar{R}\,\alpha^{*}\beta\gamma\delta^{*} & 
\bar{D}\bar{R}|\beta\gamma|^{2}+D^2\bar{R}^2|\beta\delta|^{2} & 0 \\
0 & 0 & 0 & 
D\bar{D}\bar{R}|\beta\delta|^{2}
\end{pmatrix},
\label{eq:rho_AD_damping_rev_psi}
\end{equation}
with the corresponding probability
\begin{equation}
B_{\Psi}^{(D,R)}
=\frac{
(D\bar{R}^2+\bar{D}\bar{R})|\alpha\delta|^{2}
+ \bar{D}\bar{R}|\beta\gamma|^{2}
+ (D^2\bar{R}^2+D\bar{D}\bar{R})|\beta\delta|^{2}
}{2 P_{AB}^{(D,R)} P_{CD}^{(D,R)}}.
\label{eq:prob_AD_damping_rev_psi}
\end{equation}
Its concurrence can then be calculated as
\begin{equation}
C_{\Psi}^{(D,R)} = 2\max\!\left[ 0,\,
\frac{
\bar{D}\bar{R}|\alpha\beta\gamma\delta|- \sqrt{D^2\bar{D}\bar{R}^3}\,|\alpha\delta\,\beta\delta|
}{\mathrm{Tr}[\tilde{\rho}_{\Psi}^{(D,R)}]}
\right],
\label{eq:conc_1way_psi_DR_general}
\end{equation}
which becomes
\begin{equation}
C_{\Psi}^{(D,R)}=2\max\!\left[0,\,
\frac{\big[\bar{D} - D\sqrt{\bar{D}\bar{R}}\big]
}{D(1+D)\bar{R} + \bar{D}(2+D)} \right],
\label{eq:conc_1way_psi_maxent_DR}
\end{equation}
for the maximally entangled Bell pair inputs. 
The tendencies of the amount of recovered concurrence by changing $R$ for different $D$ can be observed in Fig.~\ref{fig:Conc1way}(d). Notably, $C_{\Psi}^{(D,R)}$ increases monotonically with the reversing strength $R$, and the recovered entanglement is therefore maximized in the limit $R\to 1$. 
This means that the optimal reversing strength is given by $R_{\Psi,\mathrm{opt}}^{(D)}=1$ independent of $D$, leading to the maximal recovered concurrence $C_{\Psi,\max}^{(D)}=2/(2+D)$.
However, the probability of obtaining $\Psi^{\pm}$ outcomes given in Eq.~\eqref{eq:prob_AD_damping_rev_psi} vanishes as $R\to 1$, implying that the number of Bell pairs required to achieve the maximal recovery diverges in this limit. 
Nevertheless, by choosing a reversing strength $R<1$, substantial entanglement recovery can be achieved with a finite and experimentally feasible overhead. 
The number of required Bell pairs can be estimated for a given $D$ and $R$ by 
{\begin{equation}
Q_{\Psi}^{(D,R)}=\frac{1}{P^{(D,R)}B_{\Psi}^{(D,R)}},
\end{equation}
from Eqs.~\eqref{eq:prob_damping_rev} and \eqref{eq:prob_AD_damping_rev_psi}.
Figure~\ref{fig:Conc1way}(e) plots the recovered concurrence and Figure~\ref{fig:Conc1way}(f) plots the cost of Bell pairs as a function of $D$ for different reversing strength $R$.
We stress that, even in the ESD regime, the entanglement can be efficiently restored with a modest Bell pair overhead as we can observe in Figs.~\ref{fig:Conc1way}(e) and (f). 
For example, when $D=0.62$, the concurrence of the distributed entangled pair can be recovered up to approximately $0.47$ by consuming about $20$ initial Bell pairs by reversing operation with $R=0.9$.

As a result, in the one-way repeater model, the reversing operation can be effectively employed to recover entanglement for both the $\Phi^{\pm}$ and $\Psi^{\pm}$ BSM outcomes at a single repeater node, with a finite cost of the initial Bell pairs.
In order to extract the maximum entanglement, the optimal strategy depends on the maximum achievable concurrence for a given damping strength $D$, as we can compare from the results in Fig.~\ref{fig:Conc1way}. Although the maximum attainable concurrence for the $\Psi^{\pm}$ outcomes exceeds that of $\Phi^{\pm}$, the number of required initial Bell pairs diverges as this maximum is approached. Consequently, under finite resource constraints, the optimal strategy is to retain the resulting states associated with the $\Phi^{\pm}$ BSM outcomes and discard $\Psi^{\pm}$ outcomes, after applying the corresponding optimal reversing operations. If the objective is to maximize the average entanglement yield, likewise we discussed in the 2-way repeater model, all BSM outcomes can be retained. The concurrence averaged over all resulting states is then given by
\begin{equation}
C_{\mathrm{ave}}^{(D,R)}=2B_{\Phi}^{(D,R)}C_{\Phi}^{(D,R)}+2B_{\Psi}^{(D,R)}C_{\Psi}^{(D,R)},
\label{eq:Cave_2way}
\end{equation}
and the corresponding Bell pair cost is determined by the inverse of the success probability of the corresponding reversing operation,
\begin{equation}
Q_{\mathrm{ave}}^{(D,R)}=\frac{2}{P_{\Phi}^{(D,R)}}.
\end{equation}

To summarize this section, Table~\ref{tab:summary_esd_reversing} summarizes the occurrence of ESD without reversing operations and the effectiveness of the reversing protocol for each repeater model and BSM outcome.


\section{Entanglement recovery against photon loss}
	\label{chap:2 node model}
\vspace{8pt}

As a paradigmatic example of a practical realization of a reversing operation against photon loss, we consider the noiseless linear amplification (NLA) protocol~\cite{Ralph2009,Blandino2012,Chrzanowski2014,Xiang2010,Zavatta2011,Blandino2015,Zhao2017,Winnel2020}. 
The goal of NLA is to amplify an unknown coherent state according to the transformation $\ket{\alpha} \rightarrow \ket{g\alpha}$, where $g>1$ denotes the amplification gain. Such a noiseless amplification cannot be implemented deterministically, as it would violate the canonical commutation relation $[\hat{a},\hat{a}^{\dagger}]=1$. NLA can thus be realized as a probabilistic operation, which can be described by the map
$\ket{\alpha}\!\bra{\alpha}\rightarrow P\ket{g\alpha}\!\bra{g\alpha}
+(1-P)\ket{0}\!\bra{0}$ where $P$ is the success probability of the amplification.

NLA offers several advantages for photonic quantum information processing, allowing us to enhance the fidelity of quantum processes~\cite{Zavatta2011}, improve transmission performance through noisy channels~\cite{Blandino2015}, and enabling a variety of other applications in quantum communication and metrology~\cite{Zhao2017}. NLA can be implemented using a scheme known as the \emph{quantum scissors}~\cite{Winnel2020}, which can be understood as a teleportation-based protocol acting on a truncated Hilbert space. Its experimental setup is illustrated in Fig.~\ref{fig:NLAscheme}. For an arbitrary input state expressed in the Fock basis, $\ket{\psi} = \sum_{n=0}^{\infty} c_n \ket{n}$,
conditioned on specific photon-detection outcomes, the output state can be written by
$\hat{T}_{1}\ket{\psi}={\cal N}(c_{0}\ket{0}+gc_{1}\ket{1})$ where $\mathcal{N}$ is a normalization factor. 
Here $g=\sqrt{\eta/(1-\eta)}$ denotes the amplification gain, and $\eta$ is the transmissivity of the lower beam splitter in the setup shown in Fig.~\ref{fig:NLAscheme}.

\begin{figure}
\includegraphics[width=1.0\linewidth]{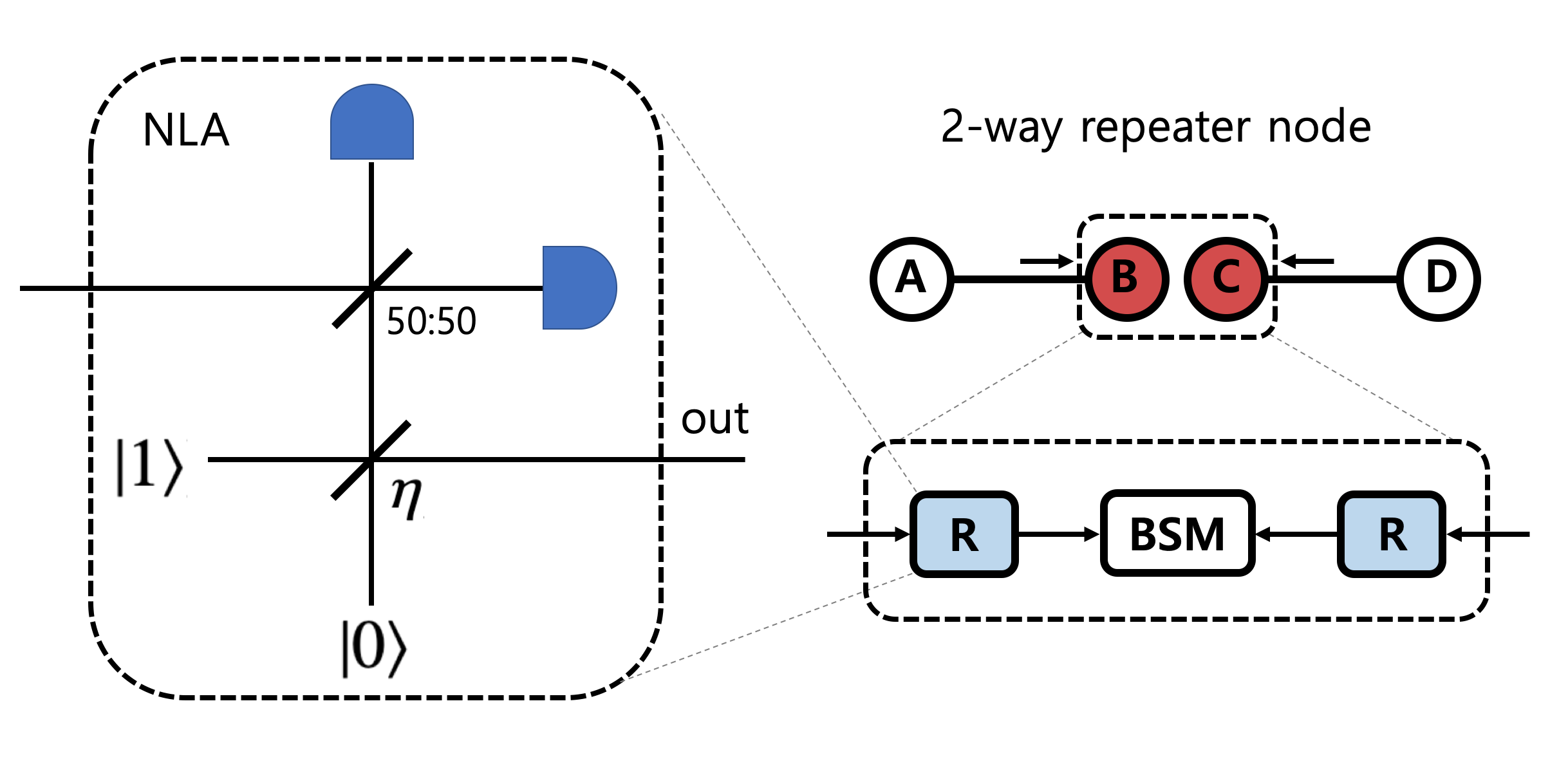}
\caption{Experimental setup for the noiseless linear amplification (NLA) based on quantum scissor. This method can be employed for reversing operation to recover the entanglement at quantum repeater nodes against photon losses.
}
\label{fig:NLAscheme}
\end{figure}
	
Within this framework, NLA provides a practical realization of the reversing operation against photon loss. Considering photonic entangled Bell pairs $\ket{\Phi^+}=(\ket{00}+\ket{11})/\sqrt{2}$ and $\ket{\Psi^+}=(\ket{01}+\ket{10})/\sqrt{2}$, which can be efficiently generated in photonic platforms~\cite{Kraus2004,Agusti2022}, photon loss in optical channels can be mapped onto an amplitude damping acting on the vacuum and single-photon basis. This establishes a one-to-one correspondence between the damping strength $D$ and the loss rate. 
So, the NLA protocol can be directly employed as a reversing operation to recover the entanglement of photonic Bell pairs. 
The effective reversing strength can be continuously tuned by adjusting the beam-splitter transmissivity $\eta \ge 1/2$ in the setup of Fig.~\ref{fig:NLAscheme}. 
More specifically, the reversing strength against photon loss is given by
$R = 1 - g^{-2} = 2 - \eta^{-1}$
which allows direct optimization of the reversing operation according to the loss rate.
We can thus apply our reversing method effectively against photon loss, and the analysis developed in the previous sections can be directly extended to optical quantum repeater architectures. 
These show that NLA-based reversal provides an effective and experimentally feasible tool to mitigate photon loss and partially recover entanglement.

\section{Discussion}
\label{chap:discussion}

We have proposed a scheme to recover the degree of distributed entanglement in noisy quantum repeaters. We introduce a reversing operation that probabilistically undoes the effect of amplitude damping decoherence, enabling heralded recovery of the entanglement without necessitating quantum memory or collective operations. We show that the degree of entanglement can be substantially recovered even under strong noise, with a success probability determined by the noise and reversing strength. Remarkably, our method enables us to recover entanglement even in the regimes where the distributed entanglement would otherwise vanish due to entanglement sudden death. We further identify optimal reversing strategies for different repeater architectures, including both two-way and one-way repeater models. We stress that substantial entanglement recovery can be achieved with a moderate Bell pair overhead. For example, it has been shown that the required Bell pair cost remains moderated numbers rather than growing rapidly below $D\lesssim0.9$ as demonstrated in Figs.~\ref{fig:Conc2way}(c) and \ref{fig:Conc1way}(c) and (f).

We have primarily considered identical amplitude damping strength $D$ for the entangled pairs in a repeater node. However, our method can be straightforwardly applied to more general situations with different damping and reversing strengths on different entangled pairs. For example, in two-way repeater model, we can calculate and analyze the resulting state by applying entanglement swapping on the pairs in Eqs.~\eqref{eq:rho_damping_rev1} and \eqref{eq:rho_damping_rev2} with different $D$ and $R$. In this case, $R$ can be locally optimized for each pair depending on $D$, while the overall qualitative behavior of entanglement recovery remains similar.
In addition, although we have focused on initial Bell pairs prepared in the state $\ket{\Phi^{+}}$, an similar analysis for $\ket{\Psi^{+}}$ can be carried out in a straightforward manner. Due to the asymmetrical nature of the amplitude damping, the effectiveness of reversing operation and the dependence on BSM outcomes differ. As demonstrated here, this asymmetry may lead to distinct robustness properties and optimal reversing strategies.


Our approach can be straightforwardly extended to quantum repeater networks with multiple intermediate nodes. The reversing operation is applied locally at each repeater node and does not require joint operations across different nodes or collective operations on multiple entangled qubits. Extending the protocol to a larger number of repeater nodes thus leads to a linear increase in resource cost of Bell pairs. On the other hand, the degree of entanglement recovered at each node is fundamentally limited by the local noise strength, implying the fact that reversing operation in each node alone cannot establish high-quality entanglement over arbitrarily long distances. This motivates the combination of our method with existing entanglement purification or distillation protocols in large-scale quantum networks.
We note that the heralded and single-copy nature of our protocol makes it well suited as a preconditioning step for entanglement purification. By probabilistically recovering entangled pairs at each repeater node, our protocol can increase the effective entanglement or fidelity of the distributed pairs before purification or distillation. 
Based on this approach, it may be possible to expand the range of distillable states, especially in the regimes where entanglement sudden death would otherwise preclude any purification. A quantitative analysis 
on the performance by combining our method with entanglement purification may be an important next step of research. 

Finally, we emphasize that our approach is experimentally feasible with currently available quantum technologies. For photonic quantum communication, photon loss can be effectively addressed in our approach by using noiseless linear amplification (NLA), which provides a tunable and heralded realization of the reversing operation. In addition, our method can be applied to mitigate the effect of amplitude damping in various platforms including photonic, ion trap and superconduing qubits~\cite{Korotkov10,Schindler13,Korotkov06,Katz08,Kim2012,Xiao2010,Lim2014,Lim2014a,Im2021,Kim2009,JCLee2011,Yan2022}. 
Our method can thus be implemented in several quantum repeater architectures without requiring substantial modifications to existing setups. We expect an experimental demonstrations of entanglement recovery based on our approach in the near future. We believe that our approach provide a practical pathway toward robust entanglement distribution in near-term quantum repeater networks.

\acknowledgments
This research was funded by National Research Foundation of Korea (RS-2022-NR068812, RS-2022-NR068814, RS-2025-00466865, RS-2025-00442855), Institute of Information \& Communications Technology Planning \& Evaluation (RS-2025-02263264), Korea Institute of Science and Technology (2E33541), and Global Partnership Program of Leading Universities in Quantum Science and Technology (RS-2025-02317602). S.B is supported by KIAS individual grant number QP100101.
	

\begin{thebibliography}{}


\bibitem{Ekert1991} A. K. Ekert, Quantum cryptography based on Bell’s
theorem, Physical Review Letters, \textbf{67}, :661–663 (1991).

\bibitem{Teleportation}
C. H. Bennett, G. Brassard, C. Crepeau, R. Jozsa, A. Peres, and W. K. Wootters, 
Teleporting an unknown quantum state via dual classical and Einstein-Podolsky-Rosen channels, Physical Review Letters {\bf 70}, 1895-1899 (1993).

\bibitem{Nielsenbook}
M. A. Nielsen and I. L. Chuang, 
Quantum Computation and Quantum Information (Cambridge Univ. Press, 2000).

\bibitem{Gisin2002}
N. Gisin, G. Ribordy, W. Tittel, and H. Zbinden, Quantum cryptography, Review of Modern Physics {\bf 74}, 145-195 (2002).

\bibitem{Kimble2008}
H. J. Kimble, \textit{The quantum internet}, Nature {\bf 453}, 1023-1030 (2008).

\bibitem{Duan2001} L.-M. Duan, M. D. Lukin, J. I. Cirac, and P. Zoller., 
Long-distance quantum communication with atomic ensembles and linear optics, 
Nature, \textbf{414}, 6862 (2001).

\bibitem{Gottesman1999} D. Gottesman and I. L. Chuang, Demonstrating the viability of universal quantum computation using teleportation and single-qubit operations. 
Nature, \textbf{402}, 6760 (1999).

\bibitem{Knill2001}
E. Knill, R. Laflamme, and G. J. Milburn. 
A scheme for efficient quantum computation with linear optics. 
Nature, \textbf{409}, 6816 (2001).

\bibitem{Lo2014}
H.-K. Lo, M. Curty, and K. Tamaki,  \textit{Secure quantum key distribution}, Nature Photonics {\bf8}, 595-604 (2014).

\bibitem{Pirandola2015} S. Pirandola, J. Eisert, C. Weedbrook, A. Furusawa, and S. L. Braunstein. 
Advances in quantum teleportation. 
Nature Photonics, \textbf{9}, 10 (2015).

\bibitem{DHKim24}
D.-H. Kim, S. Hong, Y.-S. Kim, Y. Kim, S.-W. Lee, R. C. Pooser, K. Oh, S.-Y. Lee, C. Lee, and H.-T. Lim, Distributed quantum sensing of multiple phases with fewer photons, Nat. Commun. \textbf{15}, 266 (2024).



\bibitem{Briegel1998}
H.-J. Briegel, W. D{\"u}r, J. I. Cirac, and P. Zoller.
Quantum repeaters: The role of imperfect local operations in quantum communication.
Physical Review Letters, \textbf{81}, 5932 (1998).

\bibitem{Sangouard2011}
N. Sangouard, C. Simon, H. de Riedmatten, and N. Gisin.
Quantum repeaters based on atomic ensembles and linear optics.
Reviews of Modern Physics, \textbf{83}, 33--80 (2011).

\bibitem{Meter2013} 
R. V. Meter and J. Touch. Designing quantum repeater networks. 
IEEE Communications Magazine, \textbf{51}, 64-71 (2013).

\bibitem{Wehner2018}
S. Wehner, D. Elkouss, and R. Hanson.
Quantum internet: A vision for the road ahead.
Science, \textbf{362}, eaam9288 (2018).

\bibitem{Azuma2023}
K. Azuma. S. E. Economou, D. Elkouss, P. Hilaire, L. Jiang, H.-K. Lo, and I. Tzitrin,
Quantum repeaters: from quantum networks to the quantum internet.
Review of Modern Physics, \textbf{95}, 045006 (2023).

\bibitem{Zukowski1993}
M. {\.Z}ukowski, A. Zeilinger, M. A. Horne, and A. K. Ekert.
“Event-ready-detectors” Bell experiment via entanglement swapping.
Physical Review Letters, \textbf{71}, 4287--4290 (1993).

\bibitem{Pan1998}
J.-W. Pan, D. Bouwmeester, H. Weinfurter, and A. Zeilinger.
Experimental entanglement swapping: Entangling photons that never interacted.
Physical Review Letters, \textbf{80}, 3891--3894 (1998).

\bibitem{Zurek1991}
W. H. Zurek.
Decoherence and the transition from quantum to classical.
Physics Today, \textbf{44}, 36--44 (1991).

\bibitem{Breuer2002}
H.-P. Breuer and F. Petruccione.
The Theory of Open Quantum Systems.
Oxford University Press, Oxford (2002).

\bibitem{Heinz2007}
H.-P. Breuer.
Decoherence and the transition from quantum to classical revisited.
Physical Review A, \textbf{75}, 022103 (2007).

\bibitem{Yu2009}
T. Yu and J. H. Eberly.
Sudden death of entanglement.
Science, \textbf{323}, 598--601 (2009).



\bibitem{Weinfurter94} 
H. Weinfurter, 
Experimental Bell-state analysis, 
Europhysics Letters {\bf 25}, 559 (1994).

\bibitem{Calsa2001} 
J. Calsamiglia, and N. L\"{u}tkenhaus, 
Maximum efficiency of a linear-optical Bell-state analyzer, 
Applied Physics B {\bf 72}, 67 (2001).


\bibitem{Pirandola2017}
S. Pirandola, J. Eisert, C. Weedbrook, A. Furusawa, and S. L. Braunstein.
Fundamental limits of repeaterless quantum communications.
Nature Photonics, \textbf{11}, 397--410 (2017).

\bibitem{QiChao2016} 
Q.-C. Sun, {\em et al.}
Quantum teleportation with independent sources and prior entanglement distribution over a network, 
Nature Photonics {\bf 10}, 671 (2016).

\bibitem{Valivarthi2016} 
R. Valivarthi, {\em et al.}
Quantum teleportation across a metropolitan fibre network, 
Nature Photonics {\bf 10}, 676 (2016).

\bibitem{Wzo24}
W. Zo, B Bilash, D. Lee, Y. Kim, H.-T. Lim, K. Oh, S.M. Assad, Y.-S. Kim. Entanglement swapping via lossy channels using photon-number-encoded states. Physical Review A {\bf 110}, 052603 (2024).
\bibitem{Wzo25}
W. Zo, S. Chin, Y.-S. Kim. Heralded optical entanglement distribution via lossy quantum channels: a comparative study Optics Express {\bf 33}, 12459-12474 (2025).


\bibitem{Bennett1996}
C. H. Bennett, G. Brassard, S. Popescu, B. Schumacher, J. A. Smolin, and W. K. Wootters.
Purification of noisy entanglement and faithful teleportation via noisy channels.
Physical Review Letters, \textbf{76}, 722--725 (1996).

\bibitem{Deutsch1996}
D. Deutsch, A. Ekert, R. Jozsa, C. Macchiavello, S. Popescu, and A. Sanpera.
Quantum privacy amplification and the security of quantum cryptography over noisy channels.
Physical Review Letters, \textbf{77}, 2818--2821 (1996).

\bibitem{Dur1999}
W. D{\"u}r, H.-J. Briegel, J. I. Cirac, and P. Zoller.
Quantum repeaters based on entanglement purification.
Physical Review A, \textbf{59}, 169--181 (1999).

\bibitem{Pan2001}
J.-W. Pan, C. Simon, C. Brukner, and A. Zeilinger.
Entanglement purification for quantum communication.
Nature, \textbf{410}, 1067--1070 (2001).

\bibitem{Cheong2007}
Y. W. Cheong, S.-W. Lee, J. Lee, and H.-W. Lee.
Entanglement purification for high-dimensional multipartite systems. 
Physical Review A, \textbf{76}, 042314 (2007).


\bibitem{Grice2011} 
W. P. Grice, 
Arbitrarily complete Bell-state measurement using only linear optical elements, 
Physical Review A {\bf 84}, 042331 (2011).

\bibitem{SLee13} 
S.-W. Lee and H. Jeong, 
Near-deterministic quantum teleportation and resource-efficient quantum computation using linear optics and hybrid qubits, Physical Review A {\bf 87}, 022326 (2013).

\bibitem{Ewert2014} 
F. Ewert and P. van Loock, 
3/4-efficient Bell measurement with passive linear optics and unentangled ancillae, 
Physical Review Letters {\bf 113}, 140403 (2014).

\bibitem{Lee2015}
S.-W. Lee, K. Park, T. C. Ralph, and H. Jeong, 
Nearly Deterministic Bell Measurement for Multiphoton Qubits and its Application to Quantum Information Processing,
Physical Review Letters, \textbf{114}, 113603 (2015).

\bibitem{Lee2015a} 
S.-W. Lee, K. Park, T. C. Ralph, and H. Jeong, 
Nearly deterministic Bell measurement with multiphoton entanglement for efficient quantum-information processing,
Physical Review A {\bf92}, 052324 (2015).


\bibitem{Jiang2009}
L. Jiang, J. M. Taylor, K. Nemoto, W. J. Munro, R. Van Meter, and M. D. Lukin, 
Quantum Repeater with Encoding, Physical Review A {\bf79}, 032325 (2009).

\bibitem{Munro10} 
W. J. Munro, K. A. Harrison, A. M. Stephens, S. J. Devitt, and K. Nemoto, From quantum multiplexing to high-performance quantum networking, Nature Photonics {\bf 4}, 792--796 (2010).

\bibitem{Munro12} 
W. J. Munro, A. M. Stephens, S. J. Devitt, K. A. Harrison, and K. Nemoto, Quantum communication without the necessity of quantum memories, Nature Photonics {\bf 6}, 777--781 (2012).

\bibitem{Muralidharan14} 
S. Muralidharan, J. Kim, N. L\"{u}tkenhaus, M. D. Lukin, and L. Jiang, Ultrafast and Fault-Tolerant Quantum Communication across Long Distances, Physical Review Letters {\bf 112}, 250501 (2014).

\bibitem{Muralidharan16} 
S. Muralidharan, L. Li, J. Kim, N. L\"{u}tkenhaus, M. D. Lukin, and L. Jiang, Optimal architectures for long distance quantum communication, Scientific Report {\bf 6}, 20463 (2016).

\bibitem{Azuma15} 
K. Azuma, K. Tamaki, and H.-K. Lo, 
All-photonic quantum repeaters, 
Nature Communications, \textbf{6}, 6787 (2015).

\bibitem{Ewert16} 
F. Ewert, M. Bergmann, P. van Loock, Ultrafast Long-Distance Quantum Communication with Static Linear Optics, Physical Review Letters {\bf117}, 210501 (2016).

\bibitem{Lee2019}
S.-W. Lee, T. C. Ralph, and H. Jeong, 
Fundamental building block for all-optical scalable quantum networks,
Physical Review A, \textbf{100}, 052303 (2019).

\bibitem{SML2020}
S. M. Lee, S.-W. Lee, H. Jeong, H. S. Park, Quantum teleportation of shared quantum secret, 
Physical Review Letters, \textbf{124}, 060501 (2020).



\bibitem{Jordan10} A. N. Jordan and A. N. Korotkov, 
Uncollapsing the wavefunction by undoing quantum measurements, 
Contemporary Physics {\bf 51}, 125 (2010).

\bibitem{Cheong2012} Y. W. Cheong and S.-W. Lee. Balance between information gain and reversibility in weak measurement. Physical Review Letters, \textbf{109}, 150402(2012).

\bibitem{Lim2014P}
H.-T. Lim, Y.-S. Ra, K. H. Hong, S.-W. Lee, and Y.-H. Kim.
Fundamental bounds in measurements for estimating quantum states, 
Physical Review Letters, \textbf{113}, 020504 (2014).

\bibitem{Lee2021Q}
S.-W. Lee, J. Kim, and H. Nha, 
Complete information balance in quantum measurement, 
Quantum, \textbf{5}, 414 (2021).

\bibitem{Hong2022}
S. Hong, Y.-S. Kim, Y.-W. Cho, J. Kim, S.-W. Lee, and H.-T. Lim.
Demonstration of complete information trade-off in quantum measurement, 
Physical Review Letters, \textbf{128}, 050401 (2022).

\bibitem{Lee2021R}
S.-W. Lee, D.-G. Im, Y.-H. Kim, H. Nha, and M. S. Kim,
Quantum teleportation is a reversal of quantum measurement, 
Physical Review Research, \textbf{3}, 033119 (2021).


\bibitem{Korotkov10} 
A. N. Korotkov and K. Keane, 
Decoherence suppression by quantum measurement reversal, 
Physical Review A {\bf 81}, 040103(R) (2010).

\bibitem{Kim2012}
Y.-S. Kim, J.-C. Lee, O. Kwon, and Y.-H. Kim,
Protecting entanglement from decoherence using weak measurement and quantum measurement reversal, 
Nature Physics, \textbf{8}, 117--120 (2012).

\bibitem{Xiao2010}
X. Xiao, Y. Yao, Z. Wang, and Y. Liu.
Protecting entanglement via weak measurement and quantum measurement reversal, 
Physical Review A, \textbf{82}, 032320 (2010).

\bibitem{Lim2014}
H.-T. Lim, J.-C. Lee, K.-H. Hong, and Y.-H. Kim.
Avoiding entanglement sudden death using single-qubit quantum measurement reversal, 
Optics Express, \textbf{22}, 19055--19068 (2014).

\bibitem{Lim2014a}
H.-T. Lim, J.-C. Lee, K.-H. Hong, and Y.-H. Kim, 
Observation of decoherence-induced exchange symmetry breaking in an entangled state, Phys. Rev. A \textbf{90}, 052328 (2014).

\bibitem{Im2021}
D.-G. Im, C.-H. Lee, Y. Kim, H. Nha, M. S. Kim, S.-W. Lee, and Y.-H. Kim, Fan.
Optimal teleportation via noisy quantum channels without additional qubit resources, 
npj Quantum Information, \textbf{7}, 86 (2021).


\bibitem{Kim2009}
Y.-S. Kim, Y.-W. Cho, Y.-S. Ra, Y.-H. Kim.
Reversing the weak quantum measurement for a photonic qubit, 
Optics Express, \textbf{17}, 11978--11985 (2009).

\bibitem{JCLee2011}
J.-C. Lee, Y.-C. Jeong, Y.-S. Kim, and Y.-H. Kim.
Experimental demonstration of decoherence suppression via quantum measurement reversal, 
Optics Express, \textbf{19}, 16309--16316 (2011).


\bibitem{Schindler13} P. Schindler, T. Monz, D. Nigg, J. T. Barreiro, E. A. Martinez, M. F. Brandl, M. Chwalla, M. Hennrich, and R. Blatt, Undoing a Quantum Measurement, 
Physical Review Letters, \textbf{110}, 070403 (2013).

\bibitem{Korotkov06} 
A. N. Korotkov and A. N. Jordan, 
Undoing a weak quantum measurement of a solid-state qubit, 
Physical Review Letters, \textbf{97}, 166805 (2006).

\bibitem{Katz08} 
N. Katz, M. Neeley, M. Ansmann, R. C. Bialczak, M. Hofheinz, E. Lucero, A. O'Connell, H. Wang, A. N. Cleland, J. M. Martinis, and A. N. Korotkov, 
Reversal of the weak measurement of a quantum state in a superconducting phase qubit, 
\href{https://doi.org/10.1103/PhysRevLett.101.200401}
Physical Review Letters, {\bf 101}, 200401 (2008).


\bibitem{Yan2022}
H.~Yan, Y.~Gao, Y.~Liu, Y.~Wu, M.-C.~Chen, L.~Li, Y.~Xu, H.~Deng, X.~Rong, X.~Peng, H.~Wang, A.~N.~Cleland, L.-M.~Duan, C.~P.~Sun, D.~Yu, X.~Wang, and J.-W.~Pan,
Entanglement Purification and Protection in a Superconducting Quantum Network, 
Phys. Rev. Lett. \textbf{128}, 080504 (2022).



\bibitem{Ralph2009}
T. C. Ralph and A. P. Lund.
Nondeterministic noiseless linear amplification of quantum systems, 
AIP Conference Proceedings, \textbf{1110}, 155--160 (2009).

\bibitem{Blandino2012}
R. Blandino, A. Leverrier, M. Barbieri, P. Grangier, and R. Tualle-Brouri.
Improving the maximum transmission distance of continuous-variable quantum key distribution using a noiseless amplifier,
Physical Review A, \textbf{86}, 012327 (2012).

\bibitem{Chrzanowski2014} H. M. Chrzanowski, N. Walk, S. M. Assad,J. Janousek, S. Hosseini, T. C. Ralph , T. Symul, and P. K. Lam. Measurement-based noiseless linear amplification for quantum communication, Nature Photonics, \textbf{8}, 333-338 (2014).

\bibitem{Xiang2010} G. Y. Xiang, T. C. Ralph, A. P. Lund, N. Walk, and
G. J. Pryde. Heralded noiseless linear amplification and distillation of entanglement, Nature Photonics, \textbf{4}, 316-319 (2010).
 
\bibitem{Zavatta2011} A. Zavatta, J. Fiurášek, and M. Bellini. A high-fidelity
noiseless amplifier for quantum light states, Nature Photonics, \textbf{5}, 52-56 (2011).

\bibitem{Blandino2015} R. Blandino, M. Barbieri, P. Grangier, and R. Tualle-Brouri. 
Heralded noiseless linear amplification and quantum channels, Phys. Rev. A, \textbf{91}, 062305 (2015).

\bibitem{Zhao2017} J. Zhao, J. Y. Haw, T. Symul, P. K. Lam, and S. M. Assad. Characterization of a measurement based noiseless linear amplifier and its applications. 
Phys. Rev. A, \textbf{96}, 012319 (2017).

\bibitem{Winnel2020} M. S. Winnel, N. Hosseinidehaj, and T. C. Ralph. 
Generalized quantum scissors for noiseless linear amplification, 
Phys. Rev. A, \textbf{102}, 063715 (2020).


\bibitem{Wootters1998}
W.~K.~Wootters, Entanglement of formation of an arbitrary state of two qubits,
Phys. Rev. Lett., \textbf{80}, 2245 (1998).


\bibitem{Kraus2004}
B. Kraus and J. I. Cirac, Discrete Entanglement Distribution with Squeezed Light,
Phys. Rev. Lett., \textbf{92}, 013602 (2004).

\bibitem{Agusti2022}
J. Agusti, Y. Minoguchi, J. M. Fink, and P. Rabl, Long-distance distribution of qubit-qubit entanglement using Gaussian-correlated photonic beams,
Phys. Rev. A, \textbf{105}, 062454 (2022).





\end{thebibliography}

\end{document}